\documentclass[
reprint,
superscriptaddress,
amsmath,amssymb,
aps,
longbibliography,
]{revtex4-1}

\usepackage{graphicx}
\usepackage{color}
\usepackage{dcolumn}
\usepackage{bm}
\usepackage[normalem]{ulem}

\newcommand{\ket}[1]{\left\vert{#1}\right\rangle}

\newcommand{\tone}{$T_1$}

\usepackage{textcomp}
\usepackage{xcite}
\usepackage{verbatim}

\usepackage{hyperref}
\usepackage{xr}
\makeatletter
\newcommand*{\addFileDependency}[1]{
  \typeout{(#1)}
  \@addtofilelist{#1}
  \IfFileExists{#1}{}{\typeout{No file #1.}}
}
\makeatother
\newcommand*{\myexternaldocument}[1]{
    \externaldocument{#1}
    \addFileDependency{#1.tex}
    \addFileDependency{#1.aux}
}

\myexternaldocument{supplement}
\begin{document}

	\preprint{APS/123-QED}
	
	\title{Spin readout of a CMOS quantum dot by gate reflectometry and spin-dependent tunnelling}
	
	\author{Virginia N. Ciriano-Tejel}
	\email{virginia.ciriano.17@ucl.ac.uk}
	\affiliation{London Centre for Nanotechnology, University College London, London WC1H 0AH, United Kingdom}
	\author{Michael~A.~Fogarty}
	\affiliation{London Centre for Nanotechnology, University College London, London WC1H 0AH, United Kingdom}
	\affiliation{Quantum Motion Technologies, Nexus, Discovery Way, Leeds, LS2 3AA, United Kingdom}
	\author{Simon Schaal}
	\affiliation{London Centre for Nanotechnology, University College London, London WC1H 0AH, United Kingdom}
	\affiliation{Quantum Motion Technologies, Nexus, Discovery Way, Leeds, LS2 3AA, United Kingdom}
	\author{Louis Hutin}
	\affiliation
	{CEA, LETI, Minatec Campus, F-38054 Grenoble, France}
	\author{Benoit Bertrand}
	\affiliation{CEA, LETI, Minatec Campus, F-38054 Grenoble, France}
	\author{Lisa Ibberson}
	\affiliation
	{Hitachi Cambridge Laboratory, J.J. Thomson Avenue, Cambridge CB3 0HE, United Kingdom}
	\author{M. Fernando Gonzalez-Zalba}
	\affiliation
	{Hitachi Cambridge Laboratory, J.J. Thomson Avenue, Cambridge CB3 0HE, United Kingdom}
	\author{Jing Li}
	\affiliation{Univ. Grenoble Alpes, CEA, IRIG-MEM-L\_Sim, F-38000, Grenoble, France}
	\author{Yann-Michel Niquet}
	\affiliation{Univ. Grenoble Alpes, CEA, IRIG-MEM-L\_Sim, F-38000, Grenoble, France}
	\author{Maud Vinet}
	\affiliation
	{CEA, LETI, Minatec Campus, F-38054 Grenoble, France}
	\author{John J. L. Morton}
	\email{jjl.morton@ucl.ac.uk}
	\affiliation{London Centre for Nanotechnology, University College London, London WC1H 0AH, United Kingdom}
	\affiliation{Quantum Motion Technologies, Nexus, Discovery Way, Leeds, LS2 3AA, United Kingdom}
	\affiliation
	{Dept. of Electronic \& Electrical Engineering, UCL, London WC1E 7JE, United Kingdom}

	\date{\today}
	
\begin{abstract}

Silicon spin qubits are promising candidates for realising large scale quantum processors, benefitting from a magnetically quiet host material and the prospects of leveraging the mature silicon device fabrication industry.
We report the measurement of an electron spin in a singly-occupied gate-defined quantum dot, fabricated using CMOS compatible processes at the 300~mm wafer scale. For readout, we employ spin-dependent tunneling combined with a low-footprint single-lead quantum dot charge sensor, measured using radiofrequency gate reflectometry. 
We demonstrate spin readout in two devices using this technique, obtaining valley splittings in the range 0.5--0.7~meV using excited state spectroscopy, and measure a maximum electron spin relaxation time (\tone) of $9 \pm 3$~s at 1 Tesla. These long lifetimes indicate the silicon nanowire geometry and fabrication processes employed here show a great deal of promise for qubit devices, while the spin-readout method demonstrated here is well-suited to a variety of scalable architectures.
\end{abstract}
	
	\pacs{Valid PACS appear here}
    \maketitle
 
\section{Introduction}
Spin qubits in silicon have been shown to fulfil most of the requirements to realise a quantum computer~\cite{Loss1998}, including high-fidelity qubit manipulation~\cite{Yoneda2018}, single-shot readout~\cite{Urdampilleta2019, Connors2019,Yoneda2020} and long coherence times~\cite{Veldhorst2014,Kawakami2014a}. Remaining challenges to realise a silicon quantum processor include building on recent demonstrations of two-qubit gates~\cite{ Veldhorst2015,Zajac2018, Watson2018,Xue2019} to reach the fault-tolerant threshold, as well as showing how scalable control and measurement of silicon qubits can be achieved in a way that is compatible with their high intrinsic density. 
While hole spin qubits have been demonstrated using CMOS-compatible manufacturing processes based on nanowire field effect transistors (NW-FETs)~\cite{Maurand2016}, open questions remain as to how the nanowire and its fabrication in industry standard cleanrooms impact electron spin properties such as relaxation and coherence times.

Spin qubit readout in silicon requires a spin-to-charge conversion step followed by charge detection. Various forms of spin-to-charge conversion exist such as Pauli spin blockade (PSB)~\cite{Ono2002} or spin-dependent tunnelling to a reservoir~\cite{Elzerman2004}. PSB can be detected dispersively~\cite{ West2019,  Zheng2019, House2018}, but typically charge sensors close to the qubit have been used in combination with both spin-dependent processes~\cite{ Morello2010a, Veldhorst2014, Kawakami2014a, Watson2015, Fogarty2018, Zhao2019}.  Standard three-terminal charge sensors such as the quantum point contact (QPC) or the single-electron transistor (SETs) have achieved spin readout fidelities as high as 99.9$\%$ in 6 \textmu s~\cite{ Harvey-Collard2018a, Curry2019} in DC mode and 99$\%$ in 1.6 \textmu s in RF mode~\cite{ Connors2019}. However, these sensors require two charge reservoirs near the qubit, complicating the use of this method at scale in dense qubit arrays.

As a more scalable alternative, charge sensors consisting of just two terminals in which a charge island is connected to a single reservoir, i.e. a single-electron box (SEB), have gained considerable traction~\cite{ Urdampilleta2019, House2016, Chanrion2020, Ansaloni2020}. In this method, the complex impedance of a quantum dot, which may contain both dissipative and dispersive contributions~\cite{Persson2010, Gonzalez-Zalba2015}, is measured by connecting a lumped element resonator either via a gate that controls the dot or via the reservoir. Changes in the surrounding charge environment modify the bias point of the SEB, which in turn produce an RF response conditional to the charge state of the sensed element. A spin-polarized SEB has been used to achieve spin parity readout with a fidelity of $>99\%$ in 1 ms~\cite{Urdampilleta2019}. However, a demonstration of SEB-based single spin readout is still lacking. 

In this article, we demonstrate time-averaged spin readout of a single electron in a quantum dot through spin-dependent tunnelling, detected using an adjacent quantum dot (charge sensor) which is connected to a gate-based reflectometry setup. The quantum dots are formed on opposite corners of a silicon split-gate NW-FET, fabricated using CMOS-compatible processes (Fig.~\ref{fig:Fig1}a).
We perform excited state spectroscopy of the quantum dot and measure spin relaxation times (\tone) as a function of magnetic field magnitude and orientation. We measure \tone\ up to $9 \pm 3$ seconds --- to our knowledge the longest measured so far for silicon quantum dots. This suggests that the CMOS processes and nanowire geometry do not pose limitations on spin relaxation and hold considerable promise for high-quality qubits compatible with scalable manufacture. 

\section{Setup}

\begin{figure}
\centering
\includegraphics[width=1\linewidth]{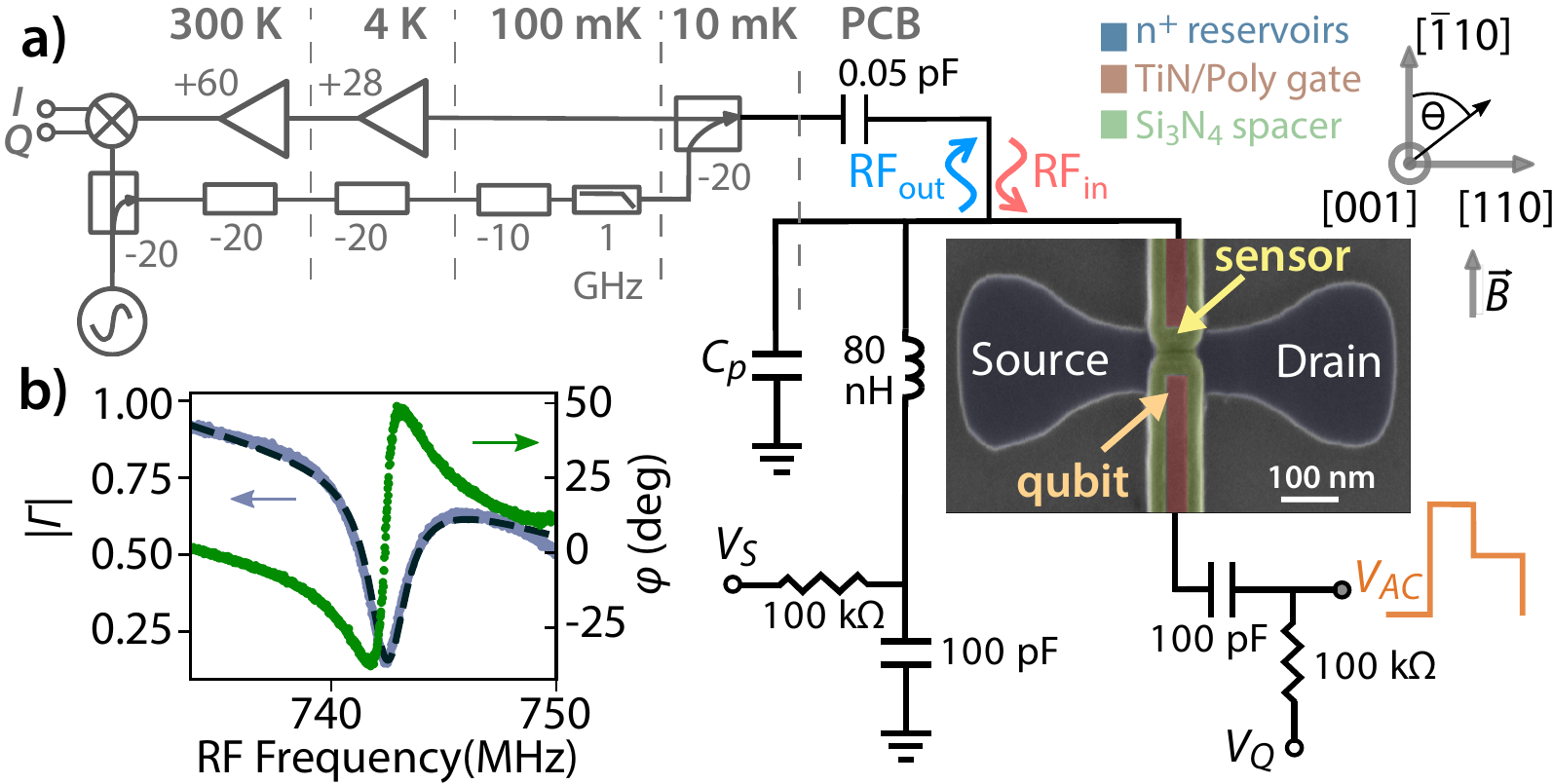}
\caption{Device and measurement setup. a) False-colour transmission electron micrograph of a silicon nanowire with a pair of split gates. Quantum dots are formed under each gate, referred to as``sensor'' and ``qubit'' dots, and controlled respectively by $V_\mathrm{S}$ and $V_\mathrm{Q}$. The sensor dot is connected to a lumped-element resonator for dispersive readout. Fast pulses, $V_\mathrm{AC}$, are applied to the qubit dot through a bias tee. To lift the spin degeneracy, a magnetic field is applied in the $[\bar{1}10]$ crystallographic direction, perpendicular to the nanowire. The magnetic field orientation can be rotated in the plane of the device, making an angle $\theta$ to $[\bar{1}10]$. b) Magnitude of the reflection coefficient, $|\Gamma|$, showing the resonator frequency at 0~T. Applying a magnetic field reduces the resonant frequency due to changes in the kinetic inductance of the superconducting inductor that forms the resonator~\cite{Lundberg2020}, see Supplementary \S II for further details.} 
    	\label{fig:Fig1}
        \end{figure}

Below, we present spin readout in two NW-FET devices, an example of which is shown in Fig.~\ref{fig:Fig1}a. ‘Device A’ has a gate length $L_\mathrm{g}=50$~nm and nanowire width $W=80$~nm, and ‘device B’ has $L_\mathrm{g}=40$~nm and $W=70$~nm. Two gates wrap onto the nanowire, in a face-to-face arrangement, with a separation between the gates, $S_\mathrm{v}$, of $50$~nm for device A and $40$~nm for device B.
Each gate can be tuned using a DC voltage to electrically induce quantum dots in the opposite corners of the silicon nanowire~\cite{Voisin2014a}, while AC signals applied to the gates are used for control and RF reflectometry read-out. The two quantum dots are tunnel-coupled (in a parallel configuration) to self-aligned, heavily implanted, n-type source and drain electron reservoirs, and capacitively coupled to each other. The device is notionally symmetric; however, we nominate one of the dots the `sensor-dot' by connecting its gate to an LC resonator for gate-based reflectometry~\cite{Ahmed2018}. 
Further details of the devices, including fabrication methods, are presented in Supplementary \S I.

\begin{figure}
\centering
\includegraphics[width=1\linewidth]{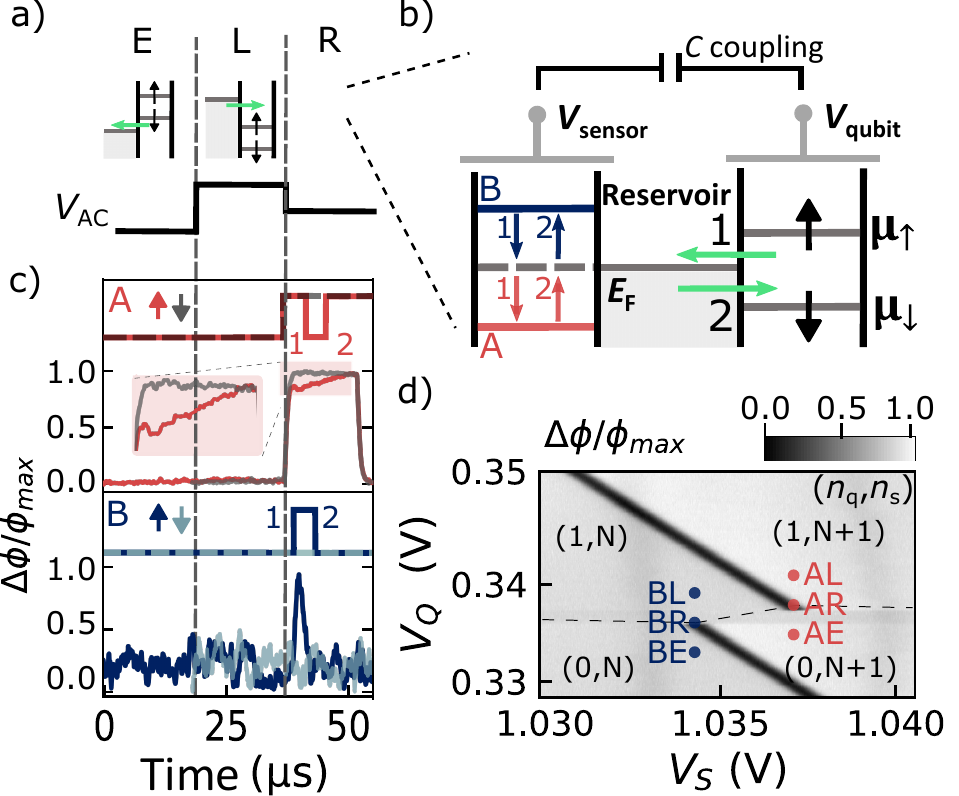}
\caption{Spin readout. a) 3-level pulse sequence applied to the qubit dot gate: first emptying the dot (E), then loading an electron with a random spin orientation into the dot (L) and finally reading the spin state (R). b) For spin readout, the qubit dot potential is tuned so its spin-up and spin-down states straddle the reservoir Fermi energy. A spin-down electron remains in the qubit dot, whereas a spin-up electron tunnels out (1) followed by a spin-down electron entering the dot (2). Due to capacitive inter-dot coupling, changes in the qubit dot charge state cause the sensor dot electrochemical potential to shift into or out of alignment with the reservoir, leading to the appearance or suppression of a phase response signal $\Delta \phi$ in reflectometry.
c) Single shot schematics and time-averaged measured phase response (1024 averages) 
for device A (red), and B (blue), where the spin up signature is respectively a dip or a peak in the phase response. d) Charge stability diagram of the double quantum dot near the $(n_q,n_s) = (1,N)\leftrightarrow(0,N+1)$ charge transition for device B (device A measurements used a nominally identical charge transition). Only the sensor dot lead-to-dot transition is visible in reflectometry. 
} 
\label{fig:Fig2}
\end{figure}

By monitoring the phase of the reflected RF signal, while the sensor and qubit potentials $V_S$ and $V_Q$ are swept, it is possible to map out charge transitions for the two quantum dots (see Fig.~\ref{fig:Fig2}d for detail and Supplementary Fig. S4 for a full stability diagram). Because the reflectometry signal is a function of the tunnelling rate of the sensor dot to the reservoir, and this rate depends on the sensor dot occupancy, $n_\mathrm{s}$, it is not straightforward to assign an electron occupation for this dot~\cite{Lundberg2020}. Nevertheless, $n_\mathrm{s}$ is not central to the charge sensing we employ here.  
The number of electrons in the qubit dot, $n_{\mathrm{q}}$, can be measured using the inter-dot capacitive coupling with the sensor: each change in $n_{\mathrm{q}}$ shifts the sensor dot electrochemical potential (see Fig.~\ref{fig:Fig2}d) allowing us to ensure complete depletion in the qubit dot by reducing $V_Q$ until no further shifts are observed by the sensor (see Fig.~S4).

\section{Spin readout}   
Once the qubit dot is depleted to its last electron, the spin degeneracy is lifted by applying a magnetic field in the plane of the device and perpendicular to the nanowire, in the $[\bar{1}10]$  crystallographic direction. The spin readout procedure follows a 3-level pulse applied to the gate forming the qubit dot, cycling between three states: ‘load’-‘empty’-‘read’ marked as ‘L’, ‘E’ and ‘R’, respectively in Fig.~\ref{fig:Fig2}a. The potential of the `read' state sits between `load' and `empty', at the $0\leftrightarrow1$ charge transition for $n_\mathrm{q}$, such that Fermi energy of the reservoir lies between the Zeeman-split spin $\ket{\uparrow}$ and $\ket{\downarrow}$ states \cite{Elzerman2004}. At this point, a spin $\ket{\downarrow}$ electron remains in the qubit dot, while a spin $\ket{\uparrow}$ electron tunnels out to the reservoir, to be subsequently replaced by a spin $\ket{\downarrow}$ electron tunnelling on the qubit dot. This spin-dependent tunnelling is detected using the sensor dot when tuned to a point in the stability diagram where the reflectometry signal depends on the qubit dot electron occupation. 
Useable `read' points in the stability diagram are ones where the $n_\mathrm{q}=0\leftrightarrow1$ charge transition intersects with the $n_\mathrm{s}=N\leftrightarrow N+1$ transition that yields a reflectometry signal. Two such points can be identified in Fig.~\ref{fig:Fig2}d labelled ‘AR’ and ‘BR’.
At `BR', a reflectometry signal (arising from the $n_\mathrm{s}=N\leftrightarrow N+1$ transition) is visible only when the qubit dot is empty ($n_q = 0$). In this case, the signature of a spin $\ket{\uparrow}$ electron on the qubit dot is the brief emergence of a reflectometry signal at the read point, as the electron tunnels out of the dot (and a new spin $\ket{\downarrow}$ tunnels in). Conversely, at `AR', a reflectometry signal is visible only when the qubit dot is occupied ($n_q = 1$), in which case the signature of spin $\ket{\uparrow}$ is a transient reduction in the signal. 
Experiments on device B used point BR for readout, while those on device A used a point equivalent to AR in the device A stability diagram. Fig.~\ref{fig:Fig2}c shows the ideal and measured spin readout traces averaged over 1024 `ELR' cycles at both `AR' and `BR'. Further tests of spin readout are shown in Supplementary \S IV.

Detecting the spin-dependent transient signals requires that the tunneling rate $\Gamma_0$ between the qubit dot and reservoir falls within the resonator bandwidth. The resonator Q-factor in our experiments was magnetic field-dependent leading to a detection bandwidth in the range 1.4--5.0~MHz.
Dot-to-reservoir tunnelling rates in these devices can be tuned by applying a voltage to a global metal top-gate (not shown in Fig.~\ref{fig:Fig1}a) or to the substrate~\cite{Ibberson2018, Ansaloni2020}. We applied 0~V and $-10$~V to the metal top gate for Devices A and B respectively, with the substrate at 0V, to achieve suitably low tunnelling rates: $\Gamma_\mathrm{0,g_A}=0.62(1)$~MHz for device A and $\Gamma_\mathrm{0,g_B}=0.97(1)$~MHz for device B. 
The spin readout signal was further optimised by fine-tuning the sensor and qubit gate voltages $V_S$ and $V_Q$. 
Through simulations of the signal dependence on these voltages, arising from the energy-dependent tunnelling rates (see Eq.~\ref{Eq:TunnelInRate} and Supplementary \S III, we obtain the tunnelling rates quoted above, as well as (for device B) an estimated g-factor of $g=1.92(11)$, and the qubit dot effective temperature of $230(9)$~mK.  This temperature limits the minimum Zeeman splitting (and hence magnetic field) at which spin readout is feasible and its elevation compared to the device temperature ($10$~mK) is attributed to the influence of the rf readout signal applied to the sensor dot acting on the qubit dot. 
The rf power used for readout can be decreased, albeit with a reduction in the spin-up visibility (see Supplementary \S VI), however, amplifiers operating at the quantum limit of introduced noise can be used to achieve higher sensitivity in RF reflectometry while using lower drive powers~\cite{Schaal2020}.

\section{Spin relaxation}
\begin{figure}
\includegraphics[width=1\linewidth]{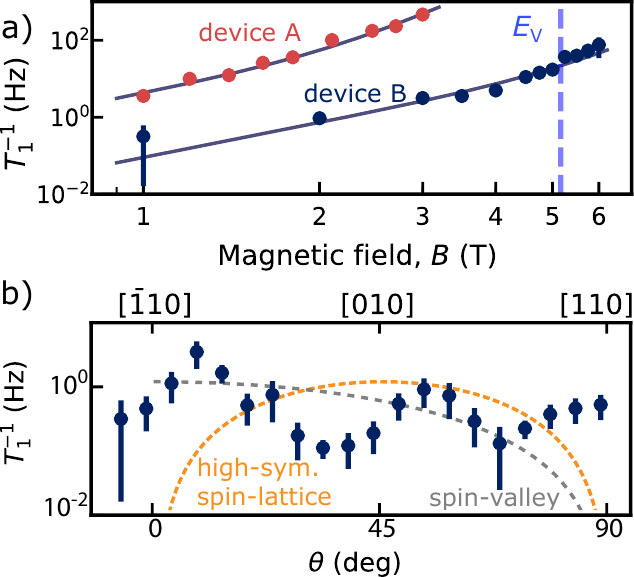}
\caption{Spin relaxation rates and their magnetic field dependence. a) Relaxation rate measured with the magnetic field applied perpendicular to the nanowire, in the plane of the device, in the $[\bar{1}10]$ crystallographic direction. Curves are fits to a general model described in the text, and $E_{\rm V}$ marks the field at which the Zeeman splitting matches the measured valley splitting in device B. b) Dependence of $T_1^{-1}$ on magnetic field orientation at 1~T for device B, where $\theta$ is the angle between the magnetic field and the $[\bar{1}10]$ crystallographic direction for device B in the nanowire plane. The angular dependence expected from spin-valley mixing in an ideal corner dot (dashed grey curve) is insufficient to explain the observed trend. Spin-lattice relaxation mechanisms  can, however, give rise to higher-order angular modulations~\cite{Glavin2003} in quantum dots with high symmetry (see for example orange dashed curve and Supplementary \S IX).
}
	\label{fig:Fig4}
    \end{figure}
    
We next consider the effect of spin relaxation by varying the duration of the `load' period in the 3-level pulse sequence. The spin of the loaded electron relaxes from its initial randomised state into the spin $\ket{\downarrow}$ ground state with a time constant \tone.
We observe exponential decays in the spin $\ket{\uparrow}$ fraction (see Supplementary \S VIII) which we fit to obtain relaxation rates $T_1^{-1}$, plotted in Fig.~\ref{fig:Fig4} as a function of magnetic field strength and orientation.
In both devices, we observe an increase in $T_1$ as the magnetic field in decreased up to a maximum of $T_1=0.28(3)$~s (device A) and $T_1=9(3)~s$ (device B) at  $B=1$~T.

The magnetic field dependence of \tone\ varies according to the relaxation mechanism and the direction of the field with respect to crystal axes. For the measurements presented in Fig.~\ref{fig:Fig4}a the magnetic field was parallel to $[\bar{1}10]$.
Spin relaxation may arise from magnetic noise at the spin Zeeman frequency or, more commonly and given some spin-orbit coupling (SOC) that mixes the spin degree of freedom with orbital or valley states, from phonon-induced electric field noise or Johnson noise.
At this field orientation, and far from any anti-crossing with higher-lying excited states~\cite{Yang2013}, the primary contributions from phonons to the relaxation rate \tone$^{-1}$ are proportional to $B^7$~\cite{Tahan2014a,Bourdet2018}, while those from Johnson Nyquist noise are proportional to $B^3$~\cite{Tahan2014a}.
We therefore fit the data in Fig.~\ref{fig:Fig4}a to a combination of such processes: $T_1^{-1}=c_\mathrm{ph} B^7+
c_\mathrm{J} B^3$ (see  Supplementary~\S IX). 

We studied the angular dependence of the spin relaxation rate in device B, rotating a 1~T field in the plane of the device. A minimum in the relaxation rate is seen as the magnetic field is parallel to the direction of the nanowire, aligned along the [110] crystallographic direction. 
Such a minimum is expected as there is no spin-valley mixing (a typically dominant spin-orbit mixing mechanism) when the magnetic field is perpendicular to a mirror symmetry plane of the device~\cite{Corna2017a, Bourdet2018}. However, we find that the usual models for spin-orbit driven relaxation~\cite{Tahan2014,Scarlino2014,Bourdet2018,Glavin2003} (see dashed lines in Fig.~\ref{fig:Fig4}b) are not able to account for all features in the angular dependence.
In general though, spin-lattice relaxation  can produce higher order harmonics in the dependence on magnetic field orientation, especially in quantum dots with high in-plane symmetry (see Supplementary \S IX). Such a high symmetry would also suggest a weak spin-valley mixing, with implications on the relaxation behaviour when then Zeeman splitting becomes comparable to the excited state valley splitting.

\section{Excited state spectroscopy}

To gain further insights into the spin relaxation mechanism for this device, we move on to study the excited valley states of this quantum dot by sweeping the voltage of the `load' stage, $V_\mathrm{Q,L}$.
The rate at which an electron loads from the reservoir into some dot state $\ket{i}$ depends on the difference in electrochemical potential, $\Delta E_\mathrm{i}$, between $\ket{i}$ and the reservoir Fermi energy. Here, we consider four dot states, $i\in\{g_\downarrow, g_\uparrow, e_\downarrow, e_\uparrow\}$, where $g$ and $e$ are respectively the ground and excited $z$-valley states, each with spin-up and spin-down states.
Assuming elastic tunnelling and a constant reservoir density of states, the loading rate follows a Fermi-Dirac distribution centred at $\Delta E_\mathrm{i}=0$, when dot and lead potentials are aligned \cite{MacLean2007, Amasha2008}:
     \begin{equation}
     \label{Eq:TunnelInRate}
        \Gamma^{\mathrm{load}}_{i}=\frac{\Gamma_{0,i}}{1+e^{\Delta E_\mathrm{i}/k_\mathrm{B} T}},
    \end{equation}
where $\Gamma_{0,i}$ is the natural tunnel rate for each state $\ket{i}$, $k_\mathrm{B}$ is the Boltzmann constant and $T$ the effective temperature. We assume here that the natural tunnel rates are spin-independent (i.e.\ for the ground states $g_\downarrow$ and $g_\uparrow$ they are equal to $\Gamma_{0,g}$, and similarly for the excited state natural tunnel rate $\Gamma_{0,e}$), as well as independent of $V_{\rm Q,L}$ over the small ($\sim1$~mV) range of voltages studied here. 
The energy separation $\Delta E_{i} $ can be tuned with $V_\mathrm{Q,L}$ as $\Delta E_\mathrm{i} = |e|\alpha_\mathrm{QQ}( V_\mathrm{i}- V_\mathrm{Q,L})$, where $V_\mathrm{i}$ is the voltage at which the dot state $\ket{i}$ and reservoir potential align and $\alpha_\mathrm{QQ}$ is the gate lever arm of the `qubit gate' to the qubit dot. 
From Eq.~\ref{Eq:TunnelInRate}, tunnelling rates tend to zero for load voltages smaller than $V_i$, and towards the natural tunnelling rate, $\Gamma_{0,i}$, for higher voltages. As a result, varying the `load' voltage $V_\mathrm{Q,L}$ changes the tunnelling rates into the various dot states, and thus the probability of loading a spin-up, which we detect using the spin-readout described above.

\begin{figure}
\includegraphics[width=1\linewidth]{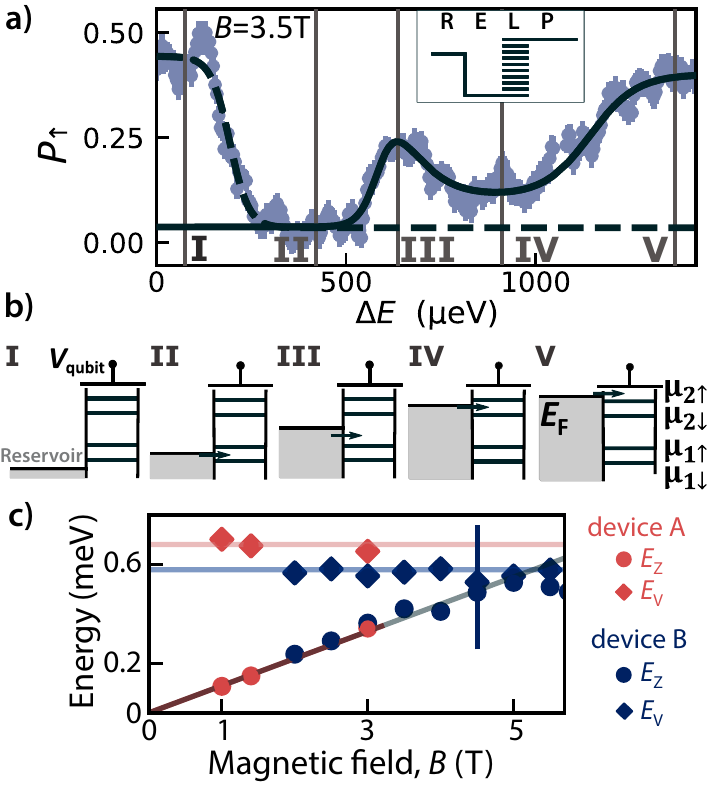}
\caption{ Excited state spectroscopy. a) Measured spin-up fraction for different load levels obtained using energy-selective loading in a  4-level pulse scheme as shown in the inset and fit. b) Illustration of different loading level regimes. (I) When the load level is too low, no electrons are loaded, and an electron with random spin tunnels in during the plunge stage. (II) If the reservoir $ E_\mathrm{F}$ is placed between the spin-up and down state, only spin-down electrons tunnel in.  (III) At higher load levels, a random spin tunnels in during the load stage. (IV) When $ E_\mathrm{F}$ lies between the spin-up and down levels of the excited state, an electron can occupy any spin state of the ground state and the spin-down excited state. Assuming fast spin-conserving relaxation from the excited to the ground state, most of electrons are found with spin-down. (V) For even higher load levels, the electron tunnels into any possible state. c) Zeeman splitting $E_\mathrm{Z}$ and excited state energy $E_\mathrm{V}$ obtained by fitting a) to Eq.~\ref{Eq:gammaup} at different magnetic fields for devices A and B. }
\label{fig:Fig3}
\end{figure}

To perform excited state spectroscopy on the qubit dot we use a 4-level pulse-sequence (`empty'-`load'-`plunge'-`read') applied to the qubit dot gate~\cite{Morello2010a}, where the additional `plunge' stage ensures that an electron is always loaded for any cycle, while the loading voltage is swept between the `empty' and `plunge' levels (see inset in Fig.~\ref{fig:Fig3}a). We define a spin-up fraction $P_{\uparrow}$ based on the integrated spin-up signal, baseline-corrected, and normalised to obtain $P_{\uparrow}=0.5$ in the limit of zero load time (to neglect relaxation) and random loading using only the `plunge' phase. The dependence of $P_{\uparrow}$ on the `load' voltage (converted to energy) is shown in Fig.~\ref{fig:Fig3}a, and can be understood by considering the schematics in Fig.~\ref{fig:Fig3}b. 
In the limit (I) of low $V_\mathrm{Q,L}$, no electron tunnels into the qubit dot during the `load' phase and an electron of random spin is loaded during `plunge'. When the Fermi energy, $E_\mathrm{F}$, of the reservoir lies between the spin-up and spin-down states (II), only spin-down electrons tunnel into the dot, and $P_{\uparrow}$ drops to zero.  
Assuming the duration of the `load' period in the pulse sequence is long compared to the natural tunnelling rates $\Gamma_{0,i}$, the transition between regions I and II is characterised by the spin-down ground state loading rate,
$\Gamma_{\mathrm{load}, g\downarrow}$:
\begin{equation}    P_\mathrm{\uparrow}=\frac{1}{2}\left(1-\frac{\Gamma^{\mathrm{load}}_{g\downarrow}}{ \Gamma_\mathrm{0,g}}\right), \label{Eq:gammadown}
\end{equation}
used to generate dashed curve in Fig.~\ref{fig:Fig3}a.

As the `load' voltage is further increased (III), both spin states can be loaded and the measured spin-up fraction increases. 
Excited states can also be measured in this way provided their decay rates to the ground state are sufficiently high~\cite{Friesen2004,Tahan2014a}. Once the spin-down excited state becomes available during the load process (IV), the measured spin-up fraction again reduces, since the excited state rapidly decays in a spin-conserving manner~\cite{Simmons2011a}. Finally, in region (V), an electron of either spin orientation can be loaded into the excited state.  In regions II--V, the measured spin up fraction can be modelled by combining all relevant rates~\cite{Simmons2011a}:

\begin{equation}
        P_{\uparrow}{(V_{\mathrm Q,L})} = \frac{
        \sum_{i=\{g\uparrow,e\uparrow\}}
        \Gamma^{\mathrm{load}}_i(V_{\mathrm Q,L})}{
        \sum_{i=\{g\uparrow,g\downarrow,e\downarrow,e\uparrow\}}
        \Gamma^{\mathrm{load}}_i(V_{\mathrm Q,L})},
        \label{Eq:gammaup}
\end{equation}
By fitting the data to Eq.~\ref{Eq:gammaup} (see solid line in Fig.\ref{fig:Fig3}a) we can extract several parameters: i) The Zeeman splitting $E_{\mathrm Z}$ between the spin-up and spin-down states (fixed to be constant for the ground and excited valley states), related to the width of regions II and IV; ii) The valley splitting $E_{\mathrm V}$, related to the separation of regions II and IV; iii) The ratio between ground and excited state natural tunnelling rates, $\Gamma_\mathrm{0}^\mathrm{e}/\Gamma_\mathrm{0}^\mathrm{g }$, related to the amplitude in region IV; and iv) the effective temperature $T$, related to the sharpness of transitions between various regions (which can be seen to be different for the ground and excited states, as discussed further in Supplementary \S VII).

Extracted values for $E_{\mathrm Z}$ and $E_{\mathrm V}$ for both devices are shown in Fig.~\ref{fig:Fig3}c as a function of magnetic field. As expected, $E_{\mathrm Z}$ shows a linear dependence with field with a g-factor of 1.91(10), while $E_{\mathrm V}$ is field-independent and measured to be 0.68(2)~meV (device A) and 0.57(3)~meV (device B). These values are broadly similar (within a factor of two) to those measured in similar nanowire  devices~\cite{Urdampilleta2019} --- furthermore, a large valley splitting is beneficial for spin qubits to remain within the computational basis states and maximise spin relaxation times~\cite{Hollmann2020}. %
The valley splitting in device B is shown as an equivalent magnetic field in Fig.~\ref{fig:Fig4}, confirming the lack of an evident relaxation `hot-spot'~\cite{Yang2013, Petit2018, Borjans2019} where $E_{\mathrm Z}\sim E_{\mathrm V}$ when there is a
finite inter-valley spin-orbit matrix element leading to spin-valley mixing. 
A possible explanation for this absence is that the corner dot has greater symmetry than expected, with two orthogonal quasi-symmetry planes, thus weakening spin-valley mixing~\cite{Corna2017a} --- this would be consistent with the complex magnetic field-orientation dependence of \tone\ discussed above. Another possible explanation is phase cancellations between the valley coupling and spin-orbit coupling matrix elements strongly suppressing spin-valley mixing ~\cite{nestoklon2006,veldhorst2015spin}.
In both cases, this interesting regime warrants investigation of further devices to ascertain the relationship between these conditions and the device geometry, growth conditions, and electrostatic environment.

\section{Conclusions and Outlook}
We have demonstrated time-averaged readout of a single spin confined in a CMOS quantum dot, using a nanowire device fabricated at the 300~mm wafer scale. 
We introduce a spin-readout method based on spin-dependent tunnelling combined with gate-based reflectometry of a neighbouring quantum dot to act as a charge sensor, representing a low-footprint approach to spin readout in silicon devices. 

Our detector bandwidth and tunnel coupling of the sensor dot to the reservoir would permit spin readout on the timescale of 10~\textmu s. However, further improvements in the signal-to-noise ratio (SNR) of the gate-based reflectometry are required to achieve high-fidelity single-shot measurements in such short times~\cite{Keith2019,Connors2019, House2016}.
For a charge transition in the sensor dot, we measure an SNR of 1 for an integration time of 50~\textmu s. 
The magnitude of the signal increases quadratically with the gate lever arm to the sensor dot~\cite{Gonzalez-Zalba2015}.
Based on the values in our device ($\alpha_{\mathrm{sensor}}=0.24$ and $\alpha_{\mathrm{qubit}}=0.47$) and similar asymmetries reported for nominally identical devices~\cite{Lundberg2020,Ibberson2020}, SNR power could be increased by 16$\times$ simply by swapping the assignment of sensor and qubit.

Further improvements in SNR power ($\sim$20$\times$ and $\sim$16$\times$ respectively) can be expected by further optimising the resonator design to detect capacitance changes~\cite{Ibberson2020} and by lowering the noise floor through use of a quantum-limited amplifier~\cite{Schaal2020}. 
Combining these methods, improvements in SNR power of three orders of magnitude are possible, bringing single-shot readout well within reach while simultaneously reducing the RF power used for readout to avoid limiting the minimum measurable Zeeman splitting.

These split-gate nanowire devices can be naturally scaled to produce 2x$n$ arrays of corner quantum dots~\cite{Hutin2019, Chanrion2020} ---  such devices could represent a 1D spin qubit array along one edge of the nanowire, where end qubits have charge sensors used for readout based on the approach presented here.
1D qubit arrays are well-suited for certain quantum simulation problems, such as a variational quantum eigensolver approach to the Hubbard model~\cite{Cai2019a, Cade2019}. Spin shuttling~\cite{Fujita2017} or qubit SWAPping~\cite{Sigillito2019} could transport qubits to the ends of the array, however, for some algorithms readout of an end-qubit ancilla is sufficient~\cite{Yuan2019}.

While it is the spin coherence time $T_2$ which ultimately limits qubit fidelity, the long spin relaxation times we measure (up to 9 s) is particularly encouraging for these devices. These indicate that both the CMOS-compatible fabrication methods and the nanowire geometry with its corner quantum dots are all consistent with large valley splittings and long spin relaxation times, making them an attractive platform for scalable quantum computing.

\begin{acknowledgments}
   We acknowledge the financial support from the European Union's Horizon 2020 research and innovation programme under grant agreement No 688539 (http://mos-quito.eu); as well as the UK's Engineering and Physical Sciences Research Council (EPSRC) through the Centre for Doctoral Training in Delivering Quantum Technologies (EP/L015242/1), QUES$^2$T (EP/N015118/1) and the Hub in Quantum Computing and Simulation
(EP/T001062/1). V.N.C.T. acknowledges the support from the Telef\'onica British-Spanish society scholarship. "Y.-M.N. and J.L. acknowledge support from the French national research agency (ANR project MAQSi). M.F.G.Z. acknowledges support from the Royal Society.
\end{acknowledgments}

\end{document}


\title{Spin readout of a CMOS quantum dot by gate reflectometry and spin-dependent tunnelling}

\author{Virginia N. Ciriano-Tejel}
	\email{virginia.ciriano.17@ucl.ac.uk}
	\affiliation{London Centre for Nanotechnology, University College London, London WC1H 0AH, United Kingdom}
	\author{Michael A. Fogarty}
	\affiliation{London Centre for Nanotechnology, University College London, London WC1H 0AH, United Kingdom}
	\affiliation{Quantum Motion Technologies Ltd, Nexus, Discovery Way, Leeds, West Yorkshire, LS2 3AA, United Kingdom}
	\author{Simon Schaal}
	\affiliation{London Centre for Nanotechnology, University College London, London WC1H 0AH, United Kingdom}
	\author{Louis Hutin}
	\affiliation
	{CEA, LETI, Minatec Campus, F-38054 Grenoble, France}
	\author{Benoit Bertrand}
	\affiliation{CEA, LETI, Minatec Campus, F-38054 Grenoble, France}
	\author{Lisa Ibberson}
	\affiliation
	{Hitachi Cambridge Laboratory, J.J. Thomson Avenue, Cambridge CB3 0HE, United Kingdom}
	\author{M. Fernando Gonzalez-Zalba}
	\affiliation
	{Hitachi Cambridge Laboratory, J.J. Thomson Avenue, Cambridge CB3 0HE, United Kingdom}
	\author{Jing Li}
	\affiliation
	{Univ. Grenoble Alpes, CEA, IRIG-MEM-L\_Sim, F-38000, Grenoble, France}
	\author{Yann-Michel Niquet}
	\affiliation
	{Univ. Grenoble Alpes, CEA, IRIG-MEM-L\_Sim, F-38000, Grenoble, France}
	\author{Maud Vinet}
	\affiliation
	{CEA, LETI, Minatec Campus, F-38054 Grenoble, France}
	\author{John J. L. Morton}
	\email{jjl.morton@ucl.ac.uk}
	\affiliation{London Centre for Nanotechnology, University College London, London WC1H 0AH, United Kingdom}
	\affiliation
	{Department of Electronic \& Electrical Engineering, University College London, London WC1E 7JE, United Kingdom}

\maketitle
		
\section{Device description}\label{SupSec_Device}

The device is a 7~nm tall silicon nanowire patterned from a silicon on insulator (SOI) substrate with a $145$-nm-thick buried oxide. An omega-shape MOS gate wraps around the Si mesa nanowire, with a stack consisting of 50~nm Poly-Si, 5~nm TiN and 6~nm thermal $\mathrm{SiO_2}$/Si.
Under a suitable gate voltage, quantum dots can form separately along the top edges of the mesa nanowire. For independent control of each dot, the wrap-around gate is split along the nanowire direction using e-beam lithography forming two gates that face each other ($G_\mathrm{sensor}$ and $G_\mathrm{qubit}$). For further control, the silicon substrate can be used as a back gate and an overarching metal line as a top gate. These two gates modify the dot electron wave function~\cite{Ibberson2018, Ansaloni2020} and, therefore, the tunnel rates between dot and reservoir. Changes in voltage applied to the metal line require stabilisation for a few days, however, thereafter the new properties remain constant and stable for extended periods sufficient for the entire experiment.

The nanowire width, $W$, and gate length, $L_\mathrm{g}$, can be engineered to achieve different inter-dot coupling and dot sizes, respectively. Device A has a gate length of $L_\mathrm{g}=50$~nm and a width of $W=80$~nm, whereas device B has $L_\mathrm{g}=40$~nm and $W=70$~nm. The splitting between gates, $S_\mathrm{v}$, is $S_\mathrm{v}=50$~nm for device A and $S_\mathrm{v}=40$~nm for device B. 

The gates are covered by 34 nm-wide $\mathrm{Si_3N_4}$ spacers. On one hand, the spacer separates the reservoirs from the central part of the intrinsic nanowire, by protecting the intrinsic silicon from the ion implantation which defines the reservoirs. And, on the other hand, it also covers the split between the independent gates, $G_\mathrm{sensor}$ and $G_\mathrm{qubit}$, since the length of both gate spacers is larger than the inter-gate gap. 

Moreover, the nanowire region below $G_\mathrm{sensor}$ is lightly Bi doped with a dose of $6\cdot
10^{10}~ $at/cm$^2$. This gives an average of approximately one Bi dopant per window of 40~nm~$\times$~40~nm.

\section{ Measurement setup}\label{SupSec_Setup} 

\subsection{Description of the DC and RF circuitry}
Measurements were performed at base temperature of a dilution refrigerator ($15$~mK). DC voltages, ($V_{\mathrm{sensor}}$, $V_{\mathrm{qubit}}$, $V_{\mathrm{top}}$), were delivered through filtered cryogenic loom. The voltage on the metal line, $V_{\mathrm{top}}$, was kept at $-10$~V in device B to reduce the qubit dot-reservoir tunnelling rate. The radio-frequency signal for gate-based readout and the fast pulses were delivered through attenuated and filtered coaxial lines. The PCB contacts are connected to the device gates through on-chip aluminium bond wires. High frequency and DC signals were combined using on-PCB bias tees. The bias tee acts on the pulses sent to the qubit gate as a high pass filter. This effect was compensated by pulse engineering using the inverse of the filter transfer function, such that after passing through the bias tee, the pulses had the desired lineshape. The resonator is formed by an $80$~nH NbN planar spiral inductor~\cite{Ahmed2018} placed in parallel to the parasitic capacitance to ground of the PCB and the device. The PCB is made from $0.8$~mm thick RO4003C with immersion silver finish.    The reflected rf signal is first amplified by 26~dB at $4$~K (LNF-LNC0.6\_2A) and further amplified at room temperature. Then, the reflected signal magnitude and phase are obtained using quadrature demodulation (Polyphase AD0540b) and measured using a digitiser (Spectrum M4i.4451-x4). %

\subsection{Resonator}
\begin{figure}
    \centering
    \includegraphics[width=1\linewidth]{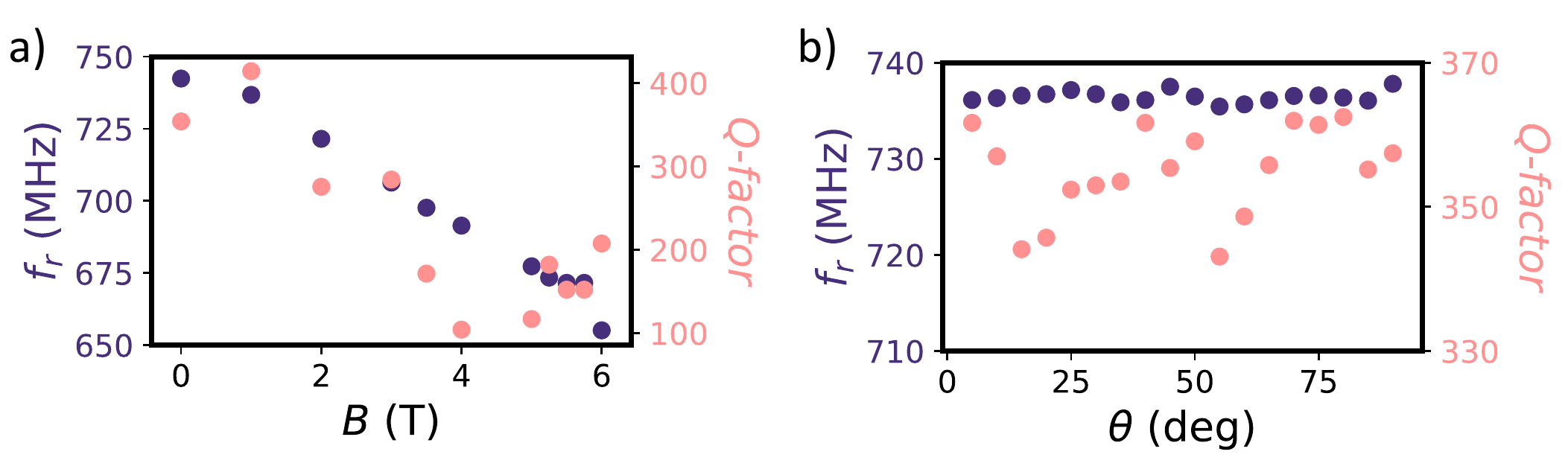}
    \caption{ a) Q-factor and resonant frequency of the resonator at different magnetic fields applied in the $[\bar{1},1,0]$ crystallographic direction. The error bars are smaller than the scatter dots. b) Q-factor and inductance when 1 Tesla was applied in different directions. $\theta$ is the angle of the magnetic field with respect to $[\bar{1},1,0]$ in the plane of the device, as in Fig.~1 of the main text.}
    \label{fig:Resonator}
\end{figure}

The resonant frequency, $f_\mathrm{r}$, and loaded Q-factor, $Q_\mathrm{L}$, vary with respect to the magnetic field due to changes in the inductor kinetic inductance (see Fig.~\ref{fig:Resonator}).  The resonant frequency can be calculated as $f_\mathrm{r}=\frac{1}{2 \pi \sqrt{L(C_\mathrm{c}+C_\mathrm{0})}}$, where $L$ is the inductor value and $C_\mathrm{0}=520$~fF is the sum of the device capacitance, $C_\mathrm{d}$, and the parasitic capacitance, $C_\mathrm{p}$. $C_c$, the coupling capacitance, is known and equal to  $C_\mathrm{c}= 50$~fF. 
The Q-factor can be extracted by measuring the $S$-parameter correspondent to the forward transmission, $S_\mathrm{21}=20\mathrm{log_{10}}{|\Gamma|}$, with a vector network analyser (VNA), which provides the reflection coefficient $|\Gamma|$ of the reflectometry setup. 
Although the resonance is undercoupled, the demodulated phase shift is still proportional to the dot capacitance shift in first order: $\Delta \phi \approx -\frac{2 Q_\mathrm{int}^3 Z_\mathrm{0}}{R_\mathrm{d}}\frac{\Delta C_\mathrm{d}}{C_\mathrm{c}+C_\mathrm{p}}$, where $Q_\mathrm{int}$ is the internal Q factor, $Z_0=50\Omega$ is the line impedance and $R_\mathrm{d}$ is the resistance seen from the dot gate.

\section{Readout offset tuning}
\label{SupSec_ReadOffsetTune}
\begin{figure}
        \centering
        \includegraphics[width=1\linewidth]{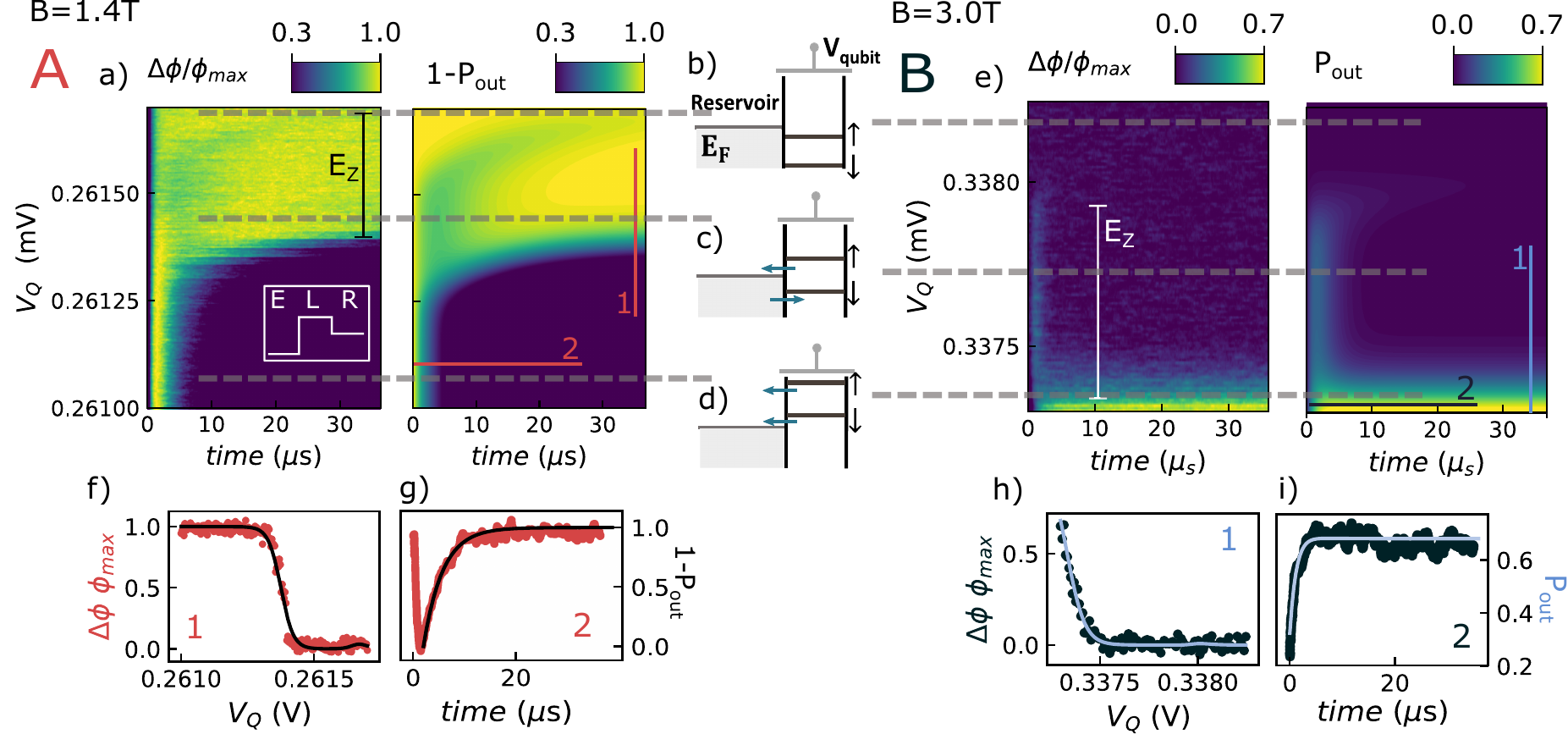}
        \caption{ a) Left: Normalised phase response of the resonator over time in the readout stage at different $V_\mathrm{Q}$ offsets for device A. The pulse sequence is depicted in the inset. Right: Simulation for an applied magnetic field of 1.4~T.  The simulation takes into account the mismatch between the bias tee cutoff frequency and the compensated pulse. b) Same for device B at B=3T. b),c) and d) diagrams of the qubit dot electrochemical potential with respect to the lead Fermi energy at three different offsets. At offset b) The electron remains in the dot. In c) only spin up electrons can tunnel out from the dot and, shortly afterwards, an electron with spin down comes back to the dot. In d)the electron always tunnels out. f) Dot occupation number along line \textit{1} comparing measurement (dotted) with simulation (line). g) Dot occupation number as a function of time at low offsets along line \textit{2}. The phase rise time due to the resonator bandwidth corresponds to the first microsecond of the graph. Panels e), h) and i) show corresponding data and simulation for device B.} 
        \label{fig:level_3}
    \end{figure}

Observing spin-dependent tunnelling requires careful tuning of the qubit gate offset voltage, $V_\mathrm{Q}$. Figs.~\ref{fig:level_3}a and \ref{fig:level_3}e show the time-dependent normalised demodulated phase at the `read' stage of the 3-level sequence, for different $V_\mathrm{Q}$, each averaged 1024 times.  For low offsets, the electron tunnels out of the qubit dot regardless of the spin state (Fig.~\ref{fig:level_3}d), whereas for higher offset voltages it always remains in the dot (Fig.~\ref{fig:level_3} b). At intermediates offsets, only electrons with spin-up can tunnel out, producing the observed spin-dependent feature. Due to the choice of different readout points in the stability diagram (see main text), the rf signal for device A is maximal when an electron is present in the qubit dot, while for device B the rf signal is maximal for the empty qubit dot.
    
    \subsection{Spin-Readout Simulations}
    
    The averaged demodulated signal during the readout stage is proportional to the expectation number of electrons in the qubit dot~\cite{Xiao2010a}. We model the qubit dot occupancy using a rate equation that considers three possible states: spin up (in the qubit dot), spin down (in the qubit dot), or no electron in the qubit dot. The ratio at which the levels are populated/emptied is given by their respective tunnel rates. The tunnelling rates depend on $V_\mathrm{Q}$, the electron temperature and the natural tunnel rate $\Gamma_\mathrm{0}$ which we take to be spin-independent. Assuming elastic tunnelling and that the reservoir has a continuous energy spectrum, the dot to reservoir tunnelling rate follows a Fermi-Dirac distribution~\cite{MacLean2007}:
    
     \begin{equation}
    \Gamma_\mathrm{in(out)}=\frac{\Gamma_0}{1+\exp\left[+(-) \Delta E/k_\mathrm{B} T\right]}.
    \end{equation}
    
    Here, $k_\mathrm{B}$ is Boltzmann's constant and $\Delta E$ is the energy difference between the relevant dot state and the lead Fermi energy. $\Delta E=|e|\alpha_\mathrm{QQ} ( V_\mathrm{\downarrow}- V_\mathrm{Q})$ for the spin down state and $\Delta E=|e|\alpha_\mathrm{QQ}( V_\mathrm{\downarrow}+E_\mathrm{z}- V_\mathrm{Q})$ for the spin up state, where  $V_\mathrm{\downarrow}$ is the voltage at which the $\ket{\downarrow}$ state and reservoir potentials align, $e$ is the electron charge, $\alpha_\mathrm{QQ}$ is the lever arm of the qubit gate on the qubit dot, and $E_\mathrm{z}$ is the Zeeman energy.   Therefore, four different tunneling rates can be defined: $\Gamma^\mathrm{in}_{\downarrow}$,$\Gamma^\mathrm{out}_{\downarrow}$,$\Gamma^\mathrm{in}_{\uparrow}$ and $\Gamma^\mathrm{out}_{\uparrow}$, i.e.\ two per dot state. 
    
    During the read stage, the sum of the probabilities of finding the electron in the dot with a spin up,  $N_{\uparrow}$, with a spin down, $N_{\downarrow}$, or out of the dot, $N_\mathrm{out}$, remains constant (and equal to one) such that the time dependent derivative of the total electronic number is equal to zero: $ \frac{ dN_\mathrm{total}}{dt}=\frac{ dN_\mathrm{out}}{dt}+ \frac{ dN_{\uparrow}}{dt}+\frac{ dN_{\downarrow}}{dt}=0$.

    The rate equation can be summarised by the following system of differential equations:
    
    \begin{equation}
    \begin{split}
       \frac{ dN_{\uparrow}}{dt} =-\Gamma^\mathrm{out}_{\uparrow} N_{\uparrow} + \Gamma^\mathrm{in}_{\uparrow} N_\mathrm{out}\\
        \frac{ dN_{\downarrow}}{dt}=-\Gamma^\mathrm{out}_{\downarrow} N_{\downarrow} + \Gamma^\mathrm{in}_{\downarrow} N_\mathrm{out}\\
        \frac{ dN_\mathrm{out}}{dt}=\Gamma^\mathrm{out}_{\uparrow} N_{\uparrow} +\Gamma^\mathrm{out}_{\downarrow} N_{\downarrow} -(\Gamma^\mathrm{in}_{\downarrow} + \Gamma^\mathrm{in}_{\uparrow})N_\mathrm{out}
    \end{split}
    \end{equation}
    
   When the system of differential equations is rewritten as a matrix, its solution has the general form:
    
    \begin{equation}
    \label{eq:model1}
    \begin{pmatrix}
    N_{\uparrow}\\
    N_{\downarrow}\\
    N_{out}\\
    \end{pmatrix}
    = xe^{v_1 t} \vec{v}_1 + ye^{v_2 t} \vec{v}_2 + ze^{v_3 t} \vec{v}_3,
    \end{equation}
    
    where $\vec{v}_\mathrm{1}$, $\vec{v}_\mathrm{2}$ and $\vec{v}_\mathrm{3}$ are the matrix eigenvectors and $v_1$, $v_2$ and $v_3$ their correspondent eigenvalues given by:

    \begin{equation}
    \centering
    \begin{split}
         v_\mathrm{1}=0\\
         v_\mathrm{2}=\frac{1}{2} (-4\Gamma_\mathrm{0} - \sqrt{(4\Gamma_\mathrm{0})^{2} - 4 (\Gamma^\mathrm{in}_{\uparrow} \Gamma^\mathrm{out}_{\downarrow} + \Gamma^\mathrm{out}_{\uparrow} (\Gamma^\mathrm{out}_{\downarrow} + \Gamma^\mathrm{in}_{\downarrow}))})\\
         v_\mathrm{3}=\frac{1}{2} (-4\Gamma_\mathrm{0} + \sqrt{(4\Gamma_\mathrm{0})^{2} - 4 (\Gamma^\mathrm{in}_{\uparrow} \Gamma^\mathrm{out}_{\downarrow} + \Gamma^\mathrm{out}_{\uparrow} (\Gamma^\mathrm{out}_{\downarrow} + \Gamma^\mathrm{in}_{\downarrow}))}).
    \end{split}
    \end{equation}
    
     $x$, $y$ and $z$ are the constants determined by the initial conditions. 
    Here, it is assumed that the qubit dot is always emptied during the empty stage and populated after the load stage such that the readout initial conditions include an electron in the dot with a random spin polarisation:
    
    \begin{equation}
    \begin{split}
         N_{\uparrow}(t=0)=1/2\\
         N_{\downarrow}(t=0)=1/2\\
         N_\mathrm{out}(t=0)=0.
    \end{split}
    \end{equation}
   
These assumptions are based on the fact that the measured tunnelling times are much shorter than the duration of the pulses. The averaged demodulated phase is proportional to the expected dot occupation number, $ 1-N_\mathrm{out}$, for device A and to $N_\mathrm{out}$ for device B.
   
Properties of the system can be obtained by examining the behaviour in particular regimes where the dynamics can be simply understood. First, at low offsets the dot state is well above the lead Fermi energy (see line-cut 2 in Fig.~\ref{fig:level_3}(a)). In this regime, $\Gamma^\mathrm{in}_{\downarrow}$ and $\Gamma^\mathrm{in}_{\uparrow}$ tend to zero, whereas $\Gamma^\mathrm{out}_{\downarrow}$ and $\Gamma^\mathrm{out}_{\uparrow}$ reach their maximum value, $\Gamma_\mathrm{0}$, 
which can thus be obtained by fitting the demodulated phase over time to an exponential decay (See Figs.~\ref{fig:level_3}(g) and \ref{fig:level_3}(i)). In this way, we obtain tunneling rates  $\Gamma_\mathrm{0,A}=0.624 \pm 0.011$~MHz for device A and $\Gamma_\mathrm{0,B}=0.970 \pm 0.012$~MHz for device B.

Second, by observing the demodulated phase with respect to $V_\mathrm{Q}$ after some time has passed (line-cut 1 in Fig.~\ref{fig:level_3}(a)), the effective temperature can be inferred. The dynamics are initially described by the complete model described in Eq.~\ref{eq:model1}, however, the effect of the negative eigenvalues fades away over time and the term $\vec{N}= x \vec{v}_1$ dominates the dot occupation. For the given initial conditions, this steady-state term reads:
\begin{equation}
         N_{\rm out}(t=\infty)=\frac{\Gamma^\mathrm{out}_{\uparrow} \Gamma^\mathrm{out}_{\downarrow}}{\Gamma^\mathrm{in}_{\uparrow} \Gamma^\mathrm{out}_{\downarrow} + \Gamma^\mathrm{out}_{\uparrow} (\Gamma^\mathrm{out}_{\downarrow} + \Gamma^\mathrm{in}_{\downarrow})},
          \label{Eq:N_out}
    \end{equation}
which for the condition $\frac{E_z}{k_BT}>>1$ simplifies to $N_{\rm out}(t=\infty)=\frac{\Gamma^\mathrm{out}_\mathrm{\downarrow}}{\Gamma_\mathrm{0}}$. Therefore, the demodulated phase with respect to $V_\mathrm{Q}$ was fitted to $1-\Gamma^\mathrm{out}_{\downarrow}/\Gamma_\mathrm{0}$ for device A and  $\Gamma^\mathrm{out}_{\downarrow}/\Gamma_\mathrm{0}$ for device B (See Figs.~\ref{fig:level_3}(f) and \ref{fig:level_3}(h)). From these fits we found an effective temperature of $0.157\pm0.012$~K for device A and $0.275\pm0.022$~K for device B (see Sec.\ref{SupSec_SigOptimisation} for full discussion of the origin of this effective temperature and noise sources).
    
The Zeeman splitting, $E_\mathrm{z}$, was calculated from the width in voltage of the spin-dependent `tail' seen in Figs.~\ref{fig:level_3}. The Zeeman splitting is plotted for different magnetic fields in Fig.~\ref{fig:anex_alpha2}.
   
To improve the fit to the data, the simulations of device A (Figs.~\ref{fig:level_3}(b) and \ref{fig:level_2}) include a voltage drift over time due to a cutoff frequency miscalculation of the bias tee high pass filter (nominally taken to be 16~kHz, but fitted to be 15.915~kHz). This small frequency mismatch does not affect measurements on the spin relaxation time.
    
\section{Two-level pulse sequence (load-read only)}\label{SupSec_GndStateDetMeas}
    
\begin{figure}
    \centering
    \includegraphics[width=1\linewidth]{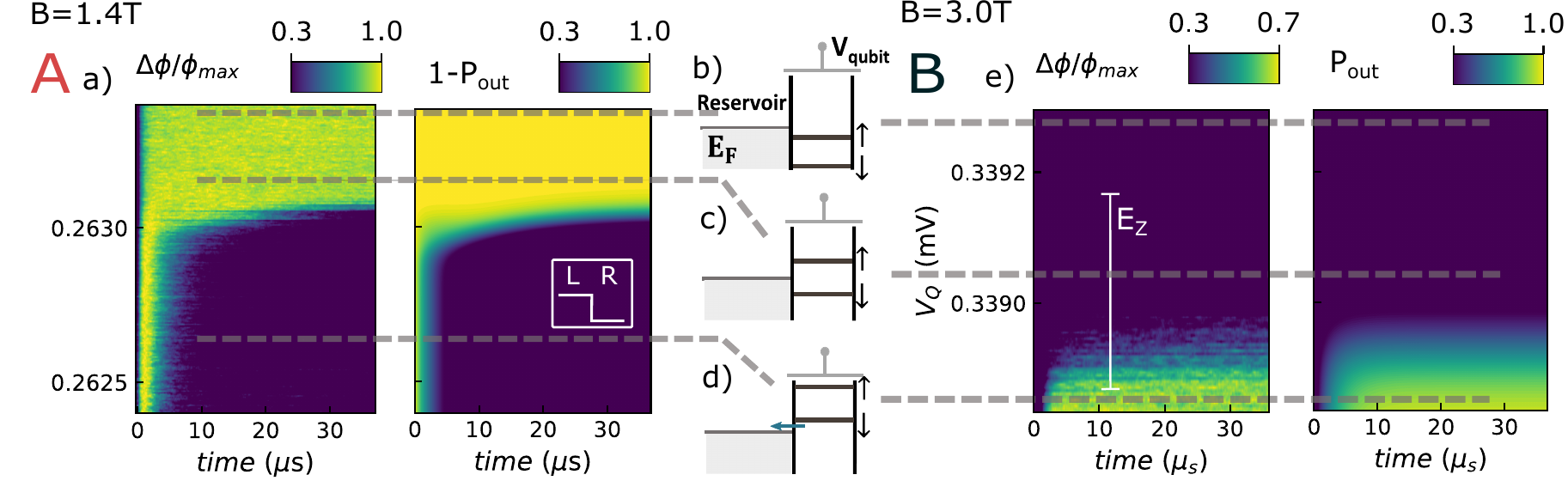}
    \caption{ Two-level pulse sequence based on `load' and `read'. a) Demodulated phase over time in the readout stage normalised at each $V_\mathrm{Q}$ offset for device A (left) and its simulation (right) for an applied magnetic field of 1.4T. The pulse sequence is depicted in the inset. b),c) and d) show diagrams of the qubit dot electrochemical potential with respect to the lead Fermi energy at three different pulse offsets, $V_\mathrm{Q}$.
    e) Is as for panel (a), but for device B at $B=3$~T.} 
        \label{fig:level_2}
    \end{figure}
    	
If the `empty' stage is removed from the pulse sequence to leave just the `load' and `read' steps, the dot remains occupied unless the electron tunnels out during the read stage. In this two-level pulse sequence, the electron in the dot eventually decays to the spin-down ground state, so that no spin up signature is observed (see Fig.~\ref{fig:level_2} for measurements and simulations).
The rate equations presented in the previous section to simulate the dot occupation number remain valid in this case --- only the initial conditions change. In the previous case, when the dot was initialised, the probability of finding the dot empty after reading out was $N_\mathrm{out}=\frac{\Gamma^\mathrm{out}_\mathrm{\downarrow}}{\Gamma_\mathrm{0}}$. For the two-level sequence, only when the dot has been emptied can a new electron be loaded (with random spin polarisation). Thus, the initial conditions are:
    \begin{equation}
    \begin{split}
         N_{\uparrow}(t=0)=\frac{\Gamma^\mathrm{out}_\mathrm{\downarrow}}{2 \Gamma_\mathrm{0}}\\
         N_{\downarrow}(t=0)=1- N_{\uparrow}(t=0)\\
         N_{out}(t=0)=0,
    \end{split}
    \end{equation}
and these are used in the simulations presented in Fig.~\ref{fig:level_2}.

\section{Gate Lever arms}\label{SupSec_LeverArms}
    
\begin{figure}
\centering
\includegraphics[width=1\linewidth]{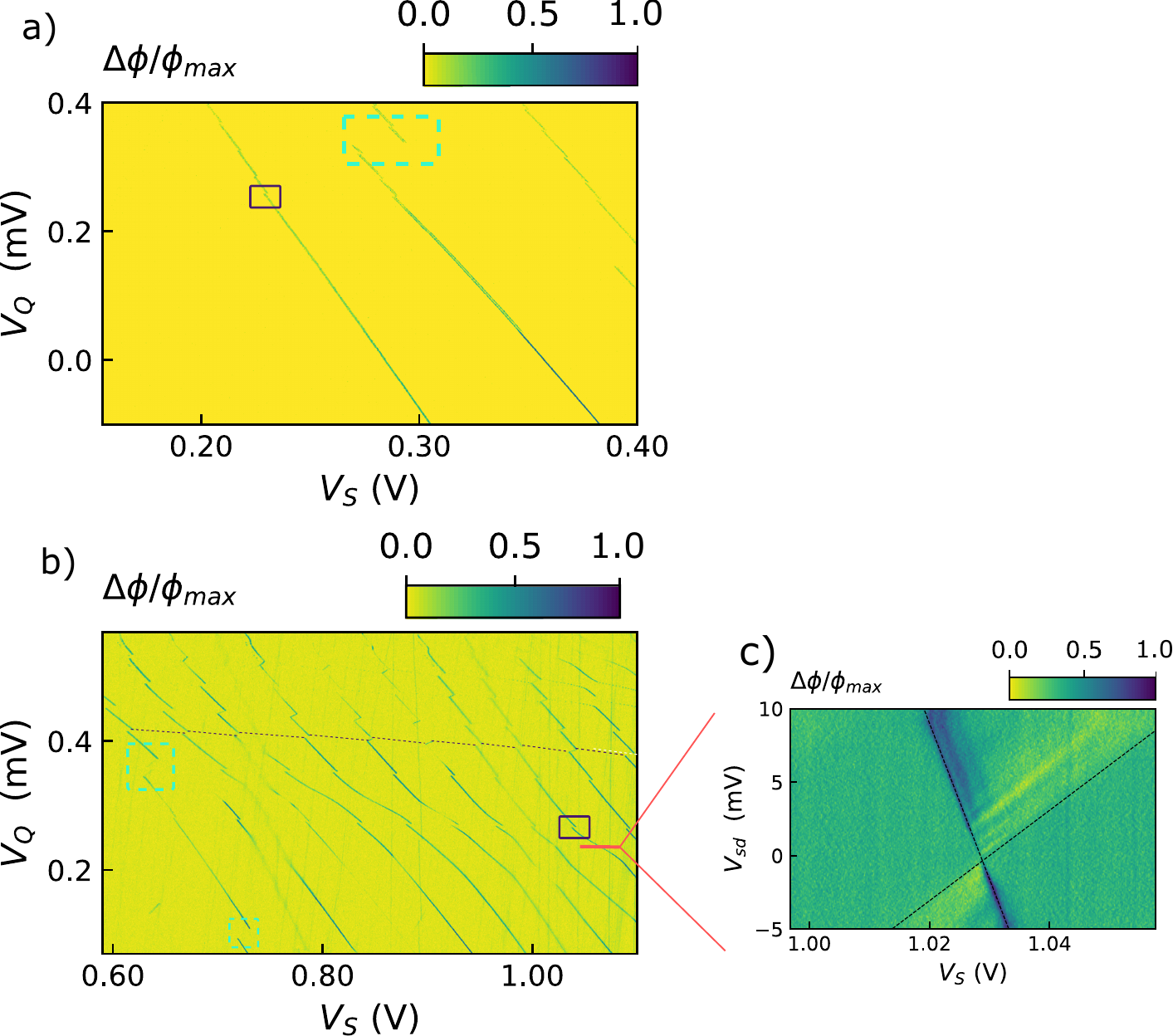}
\caption{ Lever arm and device characterisation.  a) Device A stability diagram.  A solid square indicates the readout area corresponding to the first electronic transition of the qubit dot since no other shifts are visible for a large range of smaller $V_\mathrm{Q}$. The dashed square indicates dot-donor transitions. The donor is presumed to be bismuth since the sample was bismuth-doped. b) As above for device B. The black dashed line helps the eye to follow one of the qubit dot electronic transitions. c) Coulomb diamond of the sensor dot in device B from the sensor dot-to-lead transition used in the readout. The slopes used for calculating $\alpha_\mathrm{SS}$ (see text) are marked with dashed black lines.} 
	\label{fig:anex_alpha1}
\end{figure}

The gate lever arms map the voltage applied to each gate to the electrostatic energy at the dot. In this system, where two quantum dots are placed in parallel, the Coulomb diamonds of each dot can be measured independently. The sensor dot - sensor gate lever arm ($\alpha_\mathrm{SS}$) was calculated by measuring Coulomb diamonds in the same dot-to-lead transition used for spin readout (See Fig.~\ref{fig:anex_alpha1}(c)). At $V_\mathrm{sd} \neq  0$, the transition splits in two. These two lines with slope $m_\mathrm{1}$ and $m_\mathrm{2}$, delimit the set of voltages at which the dot level is in the bias window and the lever arm can be calculated as the inverse of the slope difference: $\alpha_\mathrm{SS}=1/|1/m_\mathrm{1}-1/m_\mathrm{2}|$~\cite{Ulrich2006}.
In addition, the gate for the qubit dot can influence the sensor dot, such that in general there is a lever arm matrix~\cite{Wiel2003}:\\
	\begin{equation}	
    \begin{pmatrix}
    \Delta \mu_\mathrm{S}\\
    \Delta \mu_\mathrm{Q}\\
    \end{pmatrix}
    =
    \begin{pmatrix}
    \alpha_\mathrm{SS} & \alpha_\mathrm{SQ} \\
    \alpha_\mathrm{QS} & \alpha_\mathrm{QQ}\\
    \end{pmatrix}
    \begin{pmatrix}
    V_\mathrm{S}\\
    V_\mathrm{Q}\\
    \end{pmatrix}
	\end{equation}
               
where $\Delta \mu_\mathrm{S}$ and $\Delta \mu_\mathrm{Q}$ are the electrochemical potentials of the sensor and qubit dot, respectively.
The effect of the cross terms is visible in the stability diagram, where the slope of dot-to-lead transitions is given by the ratio between lever arms.  This way, the cross lever arm was found to be: $\alpha_\mathrm{SQ}= \frac{\Delta V _\mathrm{S}}{\Delta V_\mathrm{Q}}\cdot\alpha_\mathrm{SS}$.
	
\begin{figure}
	\centering
	\includegraphics[width=1\linewidth]{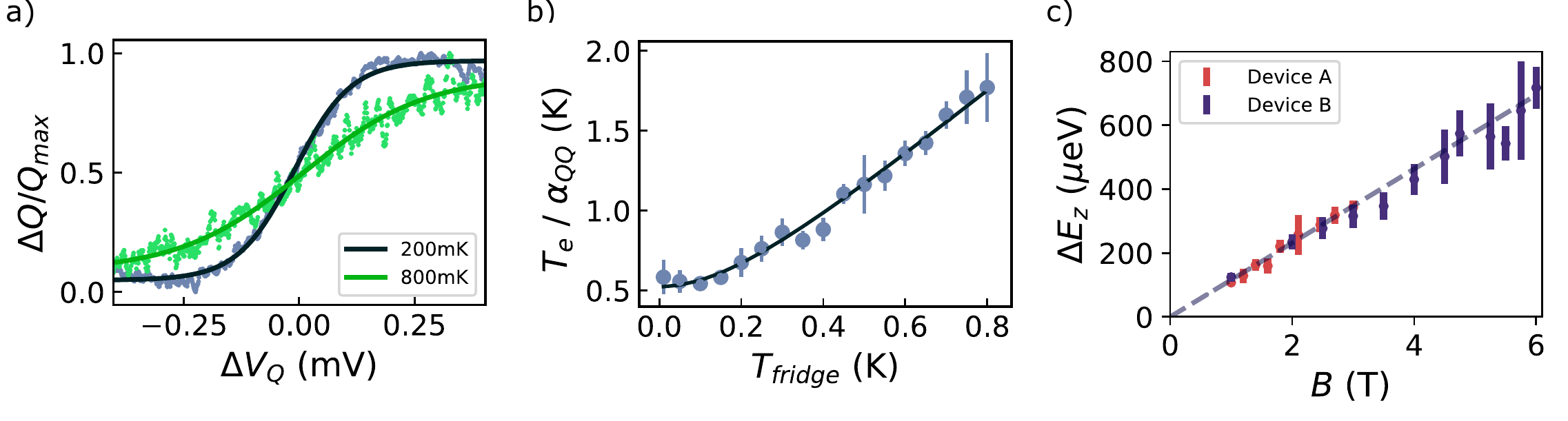}
	\caption{ Qubit dot lever arm from temperature measurements.  a)   Homodyne I/Q voltage as a function of $\Delta V_{Q}$,  at fridge temperatures of 200 and 800~mK for device B. b) Width of the Fermi-Dirac distribution measured in a) as a function of the fridge temperature and fit.  c) Zeeman energy obtained as $\Delta E_\mathrm{z}=e\alpha_\mathrm{QQ}V_\mathrm{Q}$ from Fig.~\ref{fig:level_3}a, where $\mathrm{QQ}$ is the lever arm calculated in b). Dashed line shows the Zeeman energy for $g=2$.} 
	\label{fig:anex_alpha2}
\end{figure}

The qubit dot - qubit gate lever arm, $\alpha_\mathrm{QQ}$, is determined by a temperature study in which the qubit dot occupation number is fitted with respect to $V_\mathrm{Q}$ to a Fermi distribution, as in in Figs.~\ref{fig:level_3}(g) or \ref{fig:level_3}(i). In these traces, $V_\mathrm{Q}$ is swept so the qubit dot transitions from an empty to occupied state. The broadening in this transition can have different origins: 1) the QD level broadening due to finite lifetime, 2) the effect of the rf-carrier power on the qubit dot electrochemical potential via the cross capacitance or 3) the reservoir electron temperature. We focused on the latter and an analysed this broadening with respect to the fridge temperature.  At low temperatures, the broadening is constant and, as the temperature in the fridge is raised, it increases linearly with respect to the fridge temperature (see Fig.~\ref{fig:anex_alpha2}). In this way, the temperature can be related to the transition broadening as $k_\mathrm{B}T_\mathrm{e}=e\alpha_\mathrm{QQ}V_\mathrm{Q}$ and can be fitted to $T_\mathrm{e}=\sqrt{T_\mathrm{0}^2+T_\mathrm{fridge}^2}$ ~\cite{Simmons2011a}. We obtained a $T_\mathrm{0}=230 \pm 9 $~mK and a $\alpha_\mathrm{QQ}=0.478 \pm 0.008$. 

Finally, the second cross lever arm was obtained with the stability diagram as $\alpha_\mathrm{SQ}= \alpha_\mathrm{QQ}\frac{\Delta V _\mathrm{Q}}{\Delta V_\mathrm{S}}$ using the slope of the qubit dot-to-lead transition which, although invisible, can be deduced by joining the shifts in the sensor electronic transitions (see Fig.~\ref{fig:anex_alpha1}). The lever arms values for devices A and B are summarised in Table~\ref{tab:table1}.
    
The $g$-factor was calculated using $E_\mathrm{Z}=g\mu_\mathrm{B}B=e\alpha_\mathrm{QQ} \Delta V_\mathrm{Q}$, where $\Delta V_\mathrm{Q}$ was obtained from  Figx.~\ref{fig:level_3}(a) and \ref{fig:level_3}(c) at different magnetic fields.  The calculated g-factor was $g=1.92 \pm 0.11$.

\setlength{\tabcolsep}{12pt}
 \begin{table}
  \begin{center}
    \begin{tabular}{S|r S S}
    
      \toprule 
      \multicolumn {2}{r}{ XX } & $\alpha_\mathrm{XX}$ & $\sigma_{\alpha_{\mathrm{XX}}}$\\
       \hline
       \textbf{Device A  } & \textbf{ QQ } & 0.58  & 0.03\\
       & \textbf{ SQ } & 0.070 & 0.006\\
      \hline
      \textbf{Device B  } & \textbf{ SS } &  0.239  & 0.004\\
       & \textbf{ SQ } & 0.121  & 0.002 \\
       & \textbf{ QQ } & 0.478 & 0.008\\
       & \textbf{ QS } & 0.078  & 0.004\\
      \hline 
    \end{tabular}
  \end{center}
  \caption{Lever arms. XX refers to the subindex of the alpha factor which can take the values $SS$, $SQ$, $QQ$ or $SQ$.$\sigma_{\alpha_{\mathrm{XX}}}$ refers to the standard error in the extracted values.}
    \label{tab:table1}
\end{table}

\section{Signal optimisation}\label{SupSec_SigOptimisation}

\subsection{ Power dependence of the line broadening}
To differentiate between spin states and achieve spin readout, the Zeeman splitting, $E_z=g \mu_\mathrm{B} B$ must be greater than the broadening of the $0\rightarrow1$ charge transition from in the qubit dot depicted in Fig.~\ref{fig:level_3}f. As discussed above, the broadening has at least three different sources: 1) the reservoir electron temperature, 2) the perturbations in its potential produced by the rf-carrier via cross capacitance to the sensor dot gate and 3) the dot state broadening due to tunneling. These three noise sources limit the minimum magnetic field at which the spin state can be accurately determined. 
    
Here, we study how the power from the rf-carrier sent to the `sensor' gate affects the `qubit' dot potential. Using reflectometry, the linewidth of an electronic transition is set by the electronic temperature as long as the thermal energy, $k_\mathrm{B}T$, is larger than the QD level broadening, $\hbar\gamma$, where $\gamma$ is the tunneling rate. When the electron temperature dominates, the parametric capacitance contribution due to the ability of the electron to tunnel in and out of the dot is proportional to $\Delta C_\mathrm{d} \propto \frac{1}{\cosh^2(\epsilon/2 k_\mathrm{B} T)}$, where $\epsilon$ is the quantum dot level detuning with the reservoir.  In the latter case, ($k_\mathrm{B} T\ll \mathrm{h}\gamma$), the parametric capacitance follows a Lorentzian shape: $\Delta C_\mathrm{d} \propto \frac{\hbar\gamma}{(\hbar\gamma)^2+\epsilon^2}$~\cite{Ahmed2018a}. At the same time, power applied to the sensor dot required for reflectometry can be a source for line broadening. When power broadening is dominant (See Fig. \ref{fig:Power_visibility} a), the linewidth increases as
	\begin{equation} 
	\label{eq:power}
	    \epsilon_\mathrm{\frac{1}{2}}= \epsilon_{\frac{1}{2} 0} \sqrt{1+\frac{P}{P_0}},
	\end{equation}
where $\epsilon_{\frac{1}{2}0}$ is the natural width due to electron temperature or tunnelling rates and $P_0$ is the power at which the power starts dominating the transition width. The broadening effect of the RF readout signal on the sensor potential can be translated to an effective temperature and, using the previously calculated g-factor, to an effective magnetic field, $B_{\rm noise}$ (see \S\ref{SupSec_LeverArms}). For the sensor dot, the natural width, $\epsilon_{\frac{1}{2}0}$, corresponds to a tunnelling rate of $4.87\pm0.04$~GHz or an electron temperature of $130\pm 1$~mK, consistent with previous results in Si nanowires~\cite{Ahmed2018a}.
    
The RF readout tone applied to the sensor gate can be transferred to the qubit dot due to the cross capacitance between dots or a direct capacitance between the sensor gate and the qubit dot: $\mu_\mathrm{Q}=\frac{\alpha_{QS}}{\alpha{SS}}\Delta \mu_\mathrm{S}$ (see \S~\ref{SupSec_LeverArms}). In Fig.~\ref{fig:Power_visibility} we compare i) the broadening measured on the sensor dot that has been converted to an expected qubit dot broadening, with ii) a direct measurement of the qubit dot broadening as in Fig.~\ref{fig:anex_alpha2}(a). We observe that increasing the RF readout power, increases the perturbation for both methods, however, they lead to a different natural width. This suggest that the predominant broadening at lower power does not come from the electron temperature, since it should be the same under both measurements, but from the tunneling rates which are higher in the sensor dot. At higher powers, such as the one used for spin readout ($P=-83$~dBm), we can deduce that the major contribution to line broadening (and thus effective temperature) comes from the RF tone used for readout. Its contribution can be reduced by optimising the resonator so less power is needed to show a measurable phase shift~\cite{Ahmed2018} and/or using cryogenic amplifiers with lower noise temperature such as a Josephson parametric amplifier~\cite{Schaal2020} allowing operation at lower RF power due to a decreased noise level. Moreover, although the coupling capacitance between dots is necessary for this readout, the cross capacitance between the sensor gate and the quantum dot should be as small as possible. 
	
\subsection{Visibility}
\begin{figure}
\centering
\includegraphics{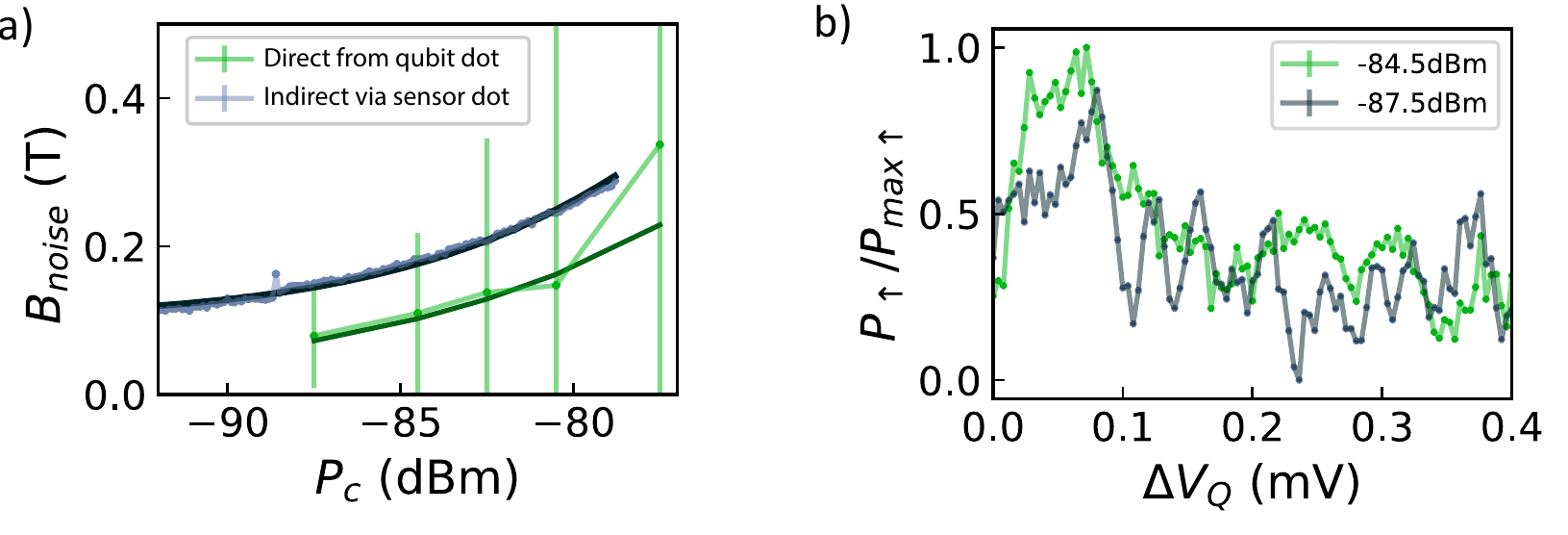}
\caption{Perturbation of the qubit dot electrochemical potential, given in Tesla, due to the rf-carrier power, $P_\mathrm{c}$. a) Blue: Electronic transition half width maximum, $\epsilon_{\frac{1}{2}}$ of the sensor dot, converted to a qubit dot potential using the cross lever arm $\alpha_\mathrm{QS}$. Error bars are smaller than the dot size. Green: A direct measurement of broadening obtained by sweeping the qubit dot voltage (see Fig.~\ref{fig:anex_alpha2}(a)).  b) Spin-up fraction at different readout offsets, $V_\mathrm{Q}$, and powers at 1~T.} 
	\label{fig:Power_visibility}
    \end{figure}

The reflectometry signal was optimised by selecting the power and qubit offset voltage that gave the highest visibility of the spin up fraction. Fig.~\ref{fig:Power_visibility}(b) shows a comparison of the spin-up fraction at different qubit dot offset voltages and RF power applied. The spin-up signature is more visible at higher voltages up to a point where the power broadening counteracts the increment in the signal.
	
\section{Excited state spectroscopy}\label{SupSec_4levelFit} 

The main text describes five different regimes of interest (I--V) in the context of excited state spectroscopy. In regimes II-V, the probability of finding an electron with spin up at different load voltages $V_\mathrm{Q,L}$ was argued to be:
     \begin{equation}
        P{\uparrow}\mathrm{(V_{Q,L})} = \frac{\Gamma_\mathrm{load,g\uparrow}+\Gamma_\mathrm{load,e\uparrow} }{\Gamma_\mathrm{load,g\uparrow}+\Gamma_\mathrm{load,e\uparrow}+\Gamma_\mathrm{load,g\downarrow}+\Gamma_\mathrm{load,e\downarrow}},
        \label{Eq:gammaup}
        \end{equation}
The parameters obtained from fitting the data to the expression above are summarised in  Table~\ref{tab:table2}. Here, $E_\mathrm{V}$ is the valley splitting, $T_\mathrm{g}$ is the effective temperature for the ground state, $T_\mathrm{e}$ is the effective temperature for the excited state and $A_\mathrm{1}=\frac{\Gamma_\mathrm{0,g}}{\Gamma_\mathrm{0,e}}$ is the ratio between the excited and the ground state natural tunneling rates.  The effective temperature of the excited states was left as a fitting parameter to include effects arising from the finite excited state lifetime. In contrast, any lifetime broadening of $\ket{g_\uparrow}$ is neglected based on the long measured T$_1$ times ($>1$~ms).

	\setlength{\tabcolsep}{12pt}
 \begin{table}
\centering

\medskip

\begin{tabular}{
  |l|l|
  r
  @{${}\pm{}$}
  l|
}
\hline
\textbf{Device A}   & $\bm{ E_{\rm V}}$ & 681 & 23 \textmu eV\\
    & $\bm{T_\mathrm{g}}$   & 370& 200 mK\\
    & $\bm{T_\mathrm{e}}$   & 510& 160 mK\\
   
        & $\bm{A_\mathrm{1}}$   & 2.0& 0.3 \\
 \hline
\textbf{Device B}  &  $\bm{E_\mathrm{V}}$  & 571& 27 \textmu eV\\
        &$\bm{T_\mathrm{g}}$   & 300& 30 mK\\
        &$\bm{T_\mathrm{e}}$   & 710& 200 mK\\
        &$\bm{A_\mathrm{1}}$   & 7.7& 0.9\\
 \hline
\end{tabular}
\caption{Fitting parameters extracted from excited state spectroscopy.}
    \label{tab:table2}
\end{table}
   
\section{$\mathit{T}_\mathrm{1}$ fitting procedure}\label{SupSec_T1Fit} 
	
\begin{figure}
	\centering
	\includegraphics[width=1\linewidth]{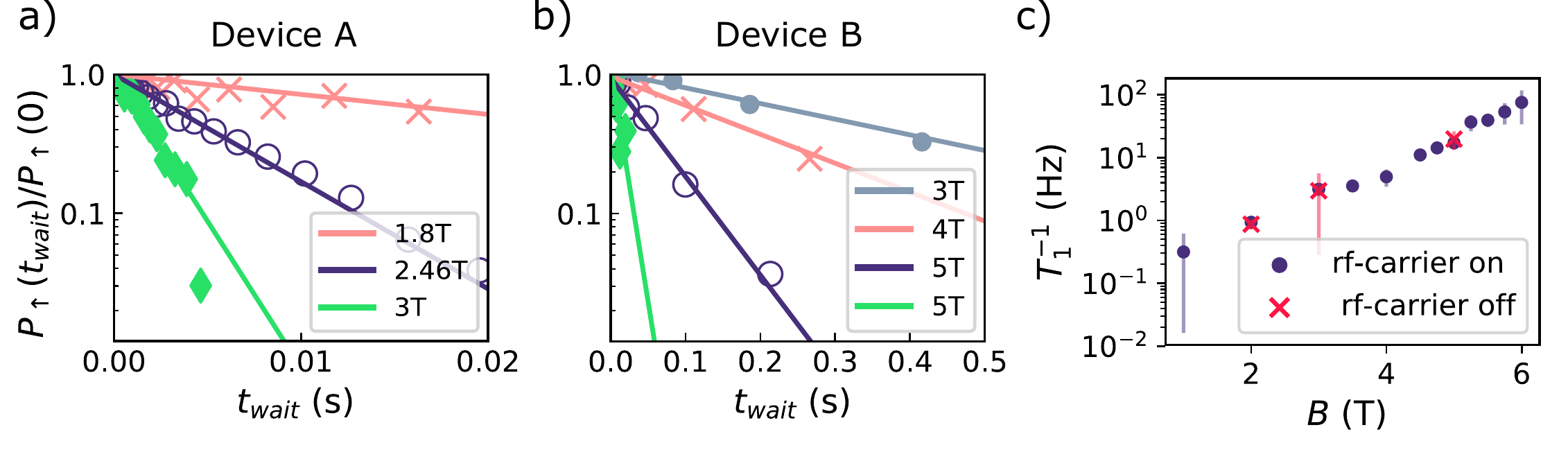}
	\caption{ a) and b) respectively show the normalised spin up fraction with respect to the waiting time for devices A and B, fitted to an exponential decay for different magnetic fields. c) Comparison between the relaxation time measured with the rf-readout tone continuously on, or  switched off during the waiting time.} 
	\label{fig:anex_T1}
    \end{figure}

The relaxation time, $\mathit{T}_\mathrm{1}$, describes the time constant at which the spin-up decays to the spin-down state. Once an electron is loaded to the dot, the probability of finding a spin up state decreases exponentially with respect to the time waited before reading its state following $\frac{P_\uparrow (t_\mathrm{wait})}{P_\uparrow (0)}=e^{-(t_\mathrm{wait}/T_\mathrm{1})}$. 
Figs.~\ref{fig:anex_T1}(a) and \ref{fig:anex_T1}(b) show the exponential fitting for several magnetic fields in device A and device B.
We also test whether the rf-carrier does not affect the relaxation during the `wait' period. Fig.~\ref{fig:anex_T1}c shows a comparison of the relaxation times measured in device B when the RF readout tone remains on throughout the 3-level pulse (purple) versus switching off the readout signal during the waiting time (red).

\section{Dependence of Spin relaxation ($T_1$) on magnetic field}\label{SupSec_T1dependence}
\subsection{Dependence on magnetic field orientation}

We discuss here the angular dependence of the spin relaxation rate in the Z valleys of silicon. 
We consider a silicon quantum dot under a finite magnetic field $\vec{B}$. We note $\ket{n,\sigma}$ and $E_{n,\sigma}=E_{n}\pm\frac{1}{2}g_0\mu_b B\sigma$ the eigenstates and eigenenergies of the dot in the absence of spin-orbit coupling, with $\sigma=\pm 1$ the spin quantized along the magnetic field axis, $g_0$ the bare gyromagnetic factor of the electron and $\mu_b$ Bohr's magneton. In these assumptions, the orbitals $\varphi_n(\vec{r})=\langle\vec{r}|n\rangle$ can be chosen real at $\vec{B}=\vec{0}$.

In the Fermi Golden rule approximation, the relaxation rate between states $\ket{\zero}\equiv\ket{0,-1}$ and $\ket{\one}\equiv\ket{0,+1}$ is typically proportional to the squared matrix element(s) $|\bra{\one}O\ket{\zero}|^2$ of one or more observable(s) $O$~\cite{Tahan2014a,Huang2014a,Bourdet2018}. We assume that $O$ is invariant under time-reversal symmetry, and does not couple spins directly [$O$ is, e.g., a local potential $V(\vec{r})$ (Johnson-Nyquist noise), an electric dipole operator $x$, $y$, or $z$ (phonons), etc...]. There must, therefore, be a mechanism such as spin-orbit coupling (SOC) mixing spins in $\ket{\zero}$ and $\ket{\one}$ in order to achieve non-zero $\bra{\one}O\ket{\zero}$'s.

Since SOC is weak in the conduction band of silicon, we can deal with it to first order in perturbation. Let $H_{\rm so}$ be the SOC Hamiltonian. The first-order $\ket{\zero}$ and $\ket{\one}$ states read:
\begin{subequations}
\begin{align}
\ket{\tilde\zero}&=\ket{\zero}+\sum_{n\ne0}\frac{\bra{n,-1}H_{\rm so}\ket{0,-1}}{E_0-E_n}\ket{n,-1}+\sum_{n}\frac{\bra{n,+1}H_{\rm so}\ket{0,-1}}{E_0-E_n-g_0\mu_bB}\ket{n,+1} \\
\ket{\tilde\one}&=\ket{\one}+\sum_{n\ne0}\frac{\bra{n,+1}H_{\rm so}\ket{0,+1}}{E_0-E_n}\ket{n,+1}+\sum_{n}\frac{\bra{n,-1}H_{\rm so}\ket{0,+1}}{E_0-E_n+g_0\mu_bB}\ket{n,-1}\,.
\end{align}
\end{subequations}
Hence, since $O$ only couples same spins,
\begin{align}
\bra{\tilde\one}O\ket{\tilde\zero}&=\sum_{n}\frac{\bra{0,+1}O\ket{n,+1}\bra{n,+1}H_{\rm so}\ket{0,-1}}{E_0-E_n-g_0\mu_bB} \nonumber \\
&+\sum_{n}\frac{\bra{0,+1}H_{\rm so}\ket{n,-1}\bra{n,-1}O\ket{0,-1}}{E_0-E_n+g_0\mu_bB}\,.
\label{eqoneOzero1}
\end{align}
We will further develop this expression to first order in $\vec{B}$, assuming $g_0\mu_bB\ll E_1-E_0$. Neglecting the action of the vector potential on the orbital motion of the electrons in a first place, the only $B$-dependent terms are the Zeeman energies on the denominators:
\begin{equation}
\frac{1}{E_0-E_n\pm g_0\mu_bB}=\frac{1}{E_0-E_n}\mp\frac{g_0\mu_bB}{(E_0-E_n)^2}
\end{equation}
Then, making use of the time-reversal symmetry relations:
\begin{subequations}
\begin{align}
&\bra{0,+1}O\ket{n,+1}=\bra{0,-1}O\ket{n,-1}^*=\bra{n,-1}O\ket{0,-1} \\
&\bra{n,+1}H_{\rm so}\ket{0,-1}=-\bra{n,-1}H_{\rm so}\ket{0,+1}^*=-\bra{0,+1}H_{\rm so}\ket{n,-1}\,
\end{align}
\label{eqtr}
\end{subequations}
we get:
\begin{equation}
\bra{\tilde\one}O\ket{\tilde\zero}=2g_0\mu_bB\sum_{n}\frac{\bra{0,+1}O\ket{n,+1}\bra{n,+1}H_{\rm so}\ket{0,-1}}{(E_n-E_0)^2}\,.
\label{eqoneOzero2}
\end{equation}
With a SOC operator of the form $H_{\rm so}=\sum_k P_k\sigma_k$, where $P_k$ are real-space operators (e.g., velocity operators) and $\sigma_k$ are the Pauli matrices (for a spin quantized along the reference axis $z$),
\begin{equation}
\bra{\tilde\one}O\ket{\tilde\zero}=iB\left(\alpha_x\bra{+1}\sigma_x\ket{-1}+\alpha_y\bra{+1}\sigma_y\ket{-1}+\alpha_z\bra{+1}\sigma_z\ket{-1}\right)\,,
\end{equation}
where the $\alpha_i$'s depend on the orbital motion of the electrons. As expected, the matrix elements $\bra{\tilde\one}O\ket{\tilde\zero}$ are proportional to $B$, since time-reversal symmetry must be broken by the magnetic field for $O$ to couple opposite spin states.

The orbitals $\varphi_n(\vec{r})$ being real, the matrix elements of the $P_k$'s must be imaginary and those of $O$ must be real according to the time-reversal symmetry relations, Eqs. (\ref{eqtr}) (this is obvious if the $P_k$'s are linear combinations of velocity operators and $O$ is one of the examples given above). Therefore, $\alpha_x$, $\alpha_y$ and $\alpha_z$ are real, and:
\begin{equation}
\bra{\tilde\one}O\ket{\tilde\zero}=iB\bra{+1}\vec{\alpha}\cdot\vec{\sigma}\ket{-1}=iB|\vec{\alpha}|\bra{+1}\vec{\sigma}_{\vec{\hat\alpha}}\ket{-1}\,,
\end{equation}
where $\vec{\sigma}_{\vec{\hat\alpha}}$ is the spin along axis $\vec{\alpha}=(\alpha_x,\alpha_y,\alpha_z)$. Since $\ket{-1}$ and $\ket{+1}$ are defined with respect to the magnetic field axis,
$|\bra{+1}\vec{\sigma}_{\vec{\hat\alpha}}\ket{-1}|=|\sin\theta_\alpha|$, where $\theta_\alpha$ is the angle between the magnetic field and the vector $\vec{\alpha}$. Hence,
\begin{equation}
|\bra{\tilde\one}O\ket{\tilde\zero}|^2\propto\sin^2\theta_\alpha\,.
\label{eqtheta}
\end{equation}
This gives rise to the simple uniaxial dependence measured for example in Ref.~\cite{Zhang2019}. In that reference, the effects of SOC are dominated by ``spin-valley'' mixing, that is by the $n=1$ term in Eq. (\ref{eqoneOzero2}) (same orbital in the other valley). In an ideal corner dot with a $(110)$ mirror symmetry plane, $\theta_\alpha$ shall be the angle with the $[110]$ axis.

The above considerations may not, however, apply when the action of the vector potential is taken into account. Indeed, in the presence of a vector potential, time-reversal symmetry transforms $\varphi_n(\vec{B}, \vec{r})$ into $\varphi_n^*(\vec{-B}, \vec{r})$, breaking Eqs. (\ref{eqtr}) and the resulting cancellations. This is not expected to make much difference for spin-valley mixing as the ground-states of both valleys effectively behave as zero (or, more generally, identical) angular momentum states and are, therefore, weakly coupled by the vector potential. Yet the effects of the vector potential may become relevant when spin-valley mixing is not dominant. 

In order to go further, we can write Eq. (\ref{eqoneOzero1}) under the form:
\begin{equation}
\bra{\tilde\one}O\ket{\tilde\zero}=\bra{+1}H_c\ket{-1}\,,
\label{eqoneHczero}
\end{equation}
where the effective Hamiltonian $H_{\rm c}$ is:
\begin{equation}
H_c=\sum_{n,k}\left(\frac{\bra{0}O\ket{n}\bra{n}P_k\ket{0}}{E_0-E_n-g_0\mu_bB}+\frac{\bra{0}P_k\ket{n}\bra{n}O\ket{0}}{E_0-E_n+g_0\mu_bB}\right)\sigma_k \,,
\end{equation}
then expand $H_{\rm c}$ to first order in $\vec{B}$ (being understood that $E_n$, $\ket{n}$ and possibly the $P_k$ operators depend on $\vec{B}$):
\begin{equation}
H_{\rm c}=\sum_{i,j}\lambda_{ij}B_i\sigma_j\,,
\label{eqHc}
\end{equation}
where $\lambda_{ij}$ are real scalars. Symmetry considerations may put constraints on the $\lambda_{ij}$'s.

Assuming $\vec{B}=B(\cos\theta, \sin\theta, 0)$, we may then sort out the angular dependence of the matrix element $\bra{+1}H_{\rm c}\ket{-1}$. The $\ket{+1}$ and $\ket{-1}$ spin states are the eigenstates of the Zeeman Hamiltonian:
\begin{equation}
H_{\rm z}=\frac{1}{2}g_0\mu_bB(\cos\theta\sigma_x+\sin\theta\sigma_y)=\frac{1}{2}g_0\mu_bB
\begin{bmatrix}
0 & e^{-i\theta} \\
e^{i\theta} & 0
\end{bmatrix}\,
\end{equation}
Hence,
\begin{align}
\ket{-1}&=\frac{e^{i\pi/4}}{\sqrt{2}}\left(e^{-i\theta/2}\ket{\uparrow}-e^{i\theta/2}\ket{\downarrow}\right) \nonumber \\
\ket{+1}&=\frac{e^{-i\pi/4}}{\sqrt{2}}\left(e^{-i\theta/2}\ket{\uparrow}+e^{i\theta/2}\ket{\downarrow}\right)\,.
\end{align}
The above phase factors have been chosen for convenience. Then,
\begin{align}
\bra{+1}\sigma_x\ket{-1}&=+\sin\theta \nonumber \\
\bra{+1}\sigma_y\ket{-1}&=-\cos\theta \nonumber \\
\bra{+1}\sigma_z\ket{-1}&=i\,.
\end{align}
Therefore, after substitution in Eq. (\ref{eqoneHczero}) and trigonometric manipulations,
\begin{align}
\bra{+1}H_{\rm c}\ket{-1}=B(a_0+ic_1\cos\theta+is_1\sin\theta+c_2\cos 2\theta+s_2\sin 2\theta)\,,
\end{align}
where $a_0$, $c_1$, $s_1$, $c_2$ and $s_2$ are real. This matrix element does, therefore, feature $\sin n\theta$ and $\cos n\theta$ harmonics up to $n=2$ -- Hence the relaxation rate, which is $\propto|\bra{\one}H_{\rm c}\ket{\zero}|^2$, features $n=0$, $n=2$, and $n=4$ harmonics. We may thus write, in general,
\begin{equation}
\Gamma=\gamma_0+\gamma_2\cos[2(\theta-\theta^0_2)]+\gamma_4\cos[4(\theta-\theta^0_4)]\,.
\label{eqGamma}
\end{equation}
Note that the relaxation rate is invariant under the transformation $\theta\to\theta+\pi$ ($\vec{B}\to-\vec{B}$), as expected. Competing relaxation mechanisms may yield different $\gamma$'s and $\theta^0$'s; yet trigonometric relations easily show that the sum over mechanisms can always be refactored under that form.

Higher-order harmonics may result from the breakdown of one of the above assumptions [first-order developments in $B$ and $H_{\rm so}$, validity of Fermi Golden Rule (multi-phonon/photon) processes], or from extrinsic contributions. Also, the prefactors of the relaxation rates scale as a power of the Larmor frequency, $\omega$ ($\omega^3$ to $\omega^5$ for phonons, $\omega$ for Johnson Nyquist noise), which may introduce extra angular dependences through the anisotropy of the g-factors. However, the contribution of g-factors to the angular dependence of the relaxation rates is presumably very weak in silicon, as they remain usually very close to 2 whatever the orientation of the magnetic field.

Examples of pure $\cos[4(\theta-\theta^0_4)]$ dependences have for example been given in Ref.~\cite{Glavin2003} (relaxation owing to phonon-induced shear strains in a highly symmetric dot). The enumeration of possible symmetry invariants in Eq. (\ref{eqHc}) suggests that the relative weight of $n=4$ harmonics shall actually increase when the dot gets more symmetric [going, e.g., from a single mirror symmetry plane ($C_s$ group) to a double mirror symmetry plane ($C_{2v}$ group)].

In the present experiments, the angular dependence of the relaxation rate is indeed dominated by $n=2$ and $n=4$ harmonics, although significant $n=6$ and $n=8$ contributions may also be needed to reproduce the behavior around $\theta=0$. Without further knowledge about the shape of that particular dot, it remains, however, difficult to make detailed predictions. Still, the presence of strong $n = 4$ harmonics suggests, as discussed above, that the relaxation is not dominated by spin-valley mixing at $B=1$ T (nor at any field given the absence of measurable hot spot at the crossing between the ground valley spin up state and the excited valley spin down state). Both the weakness of spin-valley mixing effects and the presence of sizable $n>2$ harmonics are consistent with a dot showing high in-plane symmetry~\cite{Corna2017a}. 

\subsection{Dependence on magnetic field strength}

The angular dependence described above is consistent with a small spin-valley mixing which also leads to a weak or absent hot spot in the relaxation rate when the Zeeman splitting approaches the valley splitting. We observed that the relaxation rate increases with the magnetic field, following the predicted behaviour when far from any anti-crossing with higher-lying excited states~\cite{Yang2013}.
Spin relaxation comes primarily from the spin-orbit interaction, which couples the spin degree of freedom with electrical noise. The electrical noise can have different sources, for example, the Johnson Nyquist noise, which gives a contribution proportional to $c_{jh}B^3$~\cite{Huang2014a}. In silicon, phonons can also create an electrical disturbance by deforming the lattice inhomogeneously~\cite{Hanson2007}, which, due to the non-zero dipole matrix elements between the dot levels, leads to spin relaxation. The leading contribution to this mechanism in the crystallographic direction $[\bar{1}10]$ scales as $c_\mathrm{ph}B^7$~\cite{Tahan2014a}.

We fit the relaxation rate field dependence to  the general expression $T_1^{-1}=c_\mathrm{ph} B^7+c_\mathrm{J} B^3$, obtaining the coefficients summarised in Table.\ref{tab:table3}. The large uncertainties are due to the high correlation between the two terms.

	\setlength{\tabcolsep}{12pt}
 \begin{table}
\centering

\medskip

\begin{tabular}{
  |l|l|
  r
  @{${}\pm{}$}
  l|
}
\hline
\textbf{Device A}   & $\bm{c_\mathrm{jh}}$   & 4.1 & 0.5 Hz/$T^3$\\

        &$\bm{c_\mathrm{ph}}$    & 0.171 & 0.018  Hz/$T^7$\\

 \hline
\textbf{Device B}  &  $\bm{ c_{\rm jh}}$ & 0.089 & 0.012  Hz/$T^3$\\
    & $\bm{c_\mathrm{ph}}$   & (10 & 4) $10^{-5}$ Hz/$T^7$\\
       
 \hline
\end{tabular}
\caption{Fitting parameters for the relaxation rate magnetic field dependence.}
    \label{tab:table3}
\end{table}